\documentclass[12pt]{iopart}

\usepackage{hyperref, multirow, color, bbold, textcomp}
\usepackage[dvipsnames]{xcolor}
\usepackage{orcidlink}

\newcommand{\dual}{\,{}^*\!}  \newcommand{\avg}{\mathrm{avg}}

\usepackage{tikz, varwidth}
\usetikzlibrary{calc,positioning,shapes,shapes.geometric,shapes.misc,
  shadows,arrows,arrows.meta}

\newcommand{\blue}[1]{#1}

\usepackage[normalem]{ulem} 

\usepackage{enumitem} \newcounter{requirements}[section]
\renewcommand*{\therequirements}{\thesection.\arabic{requirements}}

\newlist{reqlist}{enumerate}{3} \setlist[reqlist]{label=(\roman*),
  ref=\therequirements(\roman*), leftmargin=*, before=\raggedright}

\newcommand{\CornellPhysics}{Department of Physics, Cornell University, Ithaca,
  New York 14853, USA} \newcommand{\CornellPhysicsId}{1}

\newcommand{\CornellLepp}{Laboratory for Elementary Particle Physics, Cornell
  University, Ithaca, New York 14853, USA} \newcommand{\CornellLeppId}{2}

\newcommand{\CornellCcaps}{Cornell Center for Astrophysics and Planetary
  Science, Cornell University, Ithaca, New York 14853, USA}
\newcommand{\CornellCcapsId}{3}

\newcommand{\Unh}{Department of Physics \& Astronomy, University of New
  Hampshire, Durham, New Hampshire 03824, USA} \newcommand{\UnhId}{4}

\newcommand{\CornellAstronomy}{Department of Astronomy, Cornell University,
  Ithaca, New York 14853, USA} \newcommand{\CornellAstronomyId}{5}

\newcommand{\Caltech}{Theoretical Astrophysics 350-17, California Institute of
  Technology, Pasadena, CA 91125, USA} \newcommand{\CaltechId}{6}

\newcommand{\Fullerton}{Nicholas and Lee Begovich Center for Gravitational-Wave
  Physics and Astronomy, California State University Fullerton, Fullerton, CA
  92834, USA} \newcommand{\FullertonId}{7}

\newcommand{\Icts}{International Centre for Theoretical Sciences, Tata Institute
  of Fundamental Research, Bangalore 560089, India} \newcommand{\IctsId}{8}

\newcommand{\Wsu}{Department of Physics \& Astronomy, Washington State
  University, Pullman, Washington 99164, USA} \newcommand{\WsuId}{9}

\newcommand{\Aei}{Max Planck Institute for Gravitational Physics (Albert
  Einstein Institute), D-14467 Potsdam, Germany} \newcommand{\AeiId}{{10}}

\begin{document}

\title[BNS mergers using discontinuous Galerkin-finite difference hybrid
method]{Binary neutron star mergers using a discontinuous Galerkin-finite
  difference hybrid method}

\author{
  Nils
  Deppe$^{\CornellPhysicsId,\CornellLeppId,\CornellCcapsId}$\orcidlink{0000-0003-4557-4115},
  Francois Foucart$^\UnhId$\orcidlink{0000-0003-4617-4738},
  Marceline S.~Bonilla$^\FullertonId$\orcidlink{0000-0003-4502-528X},
  Michael Boyle$^{\CornellCcapsId}$\orcidlink{0000-0002-5075-5116},
  Nicholas
  J.~Corso$^{\CornellAstronomyId,\CornellCcapsId}$\orcidlink{0000-0002-0088-2563},
  Matthew D.~Duez$^{\WsuId}$\orcidlink{0000-0002-0050-1783},
  Matthew Giesler$^\CornellCcapsId$\orcidlink{0000-0003-2300-893X},
  Fran\c{c}ois H\'{e}bert$^\CaltechId$\orcidlink{0000-0001-9009-6955},
  Lawrence E.~Kidder$^\CornellCcapsId$\orcidlink{0000-0001-5392-7342},
  Yoonsoo Kim$^\CaltechId$\orcidlink{0000-0002-4305-6026},
  Prayush Kumar$^\IctsId$\orcidlink{0000-0001-5523-4603},
  Isaac Legred$^\CaltechId$\orcidlink{0000-0002-9523-9617},
  Geoffrey Lovelace$^\FullertonId$\orcidlink{0000-0002-7084-1070},
  Elias R.~Most$^\CaltechId$\orcidlink{0000-0002-0491-1210},
  Jordan Moxon$^\CaltechId$\orcidlink{0000-0001-9891-8677},
  Kyle C.~Nelli$^\CaltechId$\orcidlink{0000-0003-2426-8768},
  Harald P.~Pfeiffer$^\AeiId$\orcidlink{0000-0001-9288-519X},
  Mark A.~Scheel$^\CaltechId$\orcidlink{0000-0001-6656-9134},
  Saul
  A.~Teukolsky$^{\CornellCcapsId,\CaltechId}$\orcidlink{0000-0001-9765-4526},
  William Throwe$^\CornellCcapsId$\orcidlink{0000-0001-5059-4378}, and
  Nils L.~Vu$^\CaltechId$\orcidlink{0000-0002-5767-3949}}
\address{$^\CornellPhysicsId$\CornellPhysics}
\address{$^\CornellLeppId$\CornellLepp}
\address{$^\CornellCcapsId$\CornellCcaps} \address{$^\UnhId$\Unh}
\address{$^\CornellAstronomyId$\CornellAstronomy}
\address{$^\CaltechId$\Caltech} \address{$^\FullertonId$\Fullerton}
\address{$^\IctsId$\Icts}
\address{$^\WsuId$\Wsu}
\address{$^\AeiId$\Aei}
\ead{nd357@cornell.edu}

\begin{abstract}

  We present a discontinuous Galerkin-finite difference hybrid scheme that
  allows high-order shock capturing with the discontinuous Galerkin method for
  general relativistic magnetohydrodynamics in dynamical spacetimes. We present
  several optimizations and stability improvements to our algorithm that allow
  the hybrid method to successfully simulate single, rotating, and binary
  neutron stars. The hybrid method achieves the efficiency of discontinuous
  Galerkin methods throughout almost the entire spacetime during the inspiral
  phase, while being able to robustly capture shocks and resolve the stellar
  surfaces. We also use Cauchy-Characteristic evolution to compute the first
  gravitational waveforms at future null infinity from binary neutron star
  mergers. The simulations presented here are the first successful binary
  neutron star inspiral and merger simulations using discontinuous Galerkin
  methods.

\end{abstract}


\submitto{\CQG}

\section{Introduction\label{sec:dgscl introduction}}

The discontinuous Galerkin (DG) method was first presented by Reed and
Hill~\cite{reed1973triangular} to solve the neutron transport equation. Later,
in a series of seminal papers, Cockburn and Shu applied the DG method to
nonlinear hyperbolic conservation laws~\cite{Cockburn1989tvb, COCKBURN198990,
  1990MaCom..54..545C}. An important property of DG methods is that they
guarantee linear stability in the $L_2$ norm for arbitrary high
order~\cite{jiang1994cell, barth2001energy, hou2007solutions}. While this means
the DG method is very robust, DG alone is still subject to Godunov's
theorem~\cite{Godunov1959}: at high order it produces oscillatory
solutions. This means DG requires a nonlinear supplemental method for stability
in the presence of discontinuities and large gradients. We extend the
discontinuous Galerkin-finite difference (DG-FD) hybrid method developed
in~\cite{Deppe:2021ada} to dynamical spacetimes. The method is implemented in
the open-source numerical relativity code,
\texttt{SpECTRE}\cite{spectrecode_2024}.

Spectral-type methods have proven extremely useful in producing a large number
of long and accurate gravitational waveforms from binary black hole merger
simulations~\cite{Boyle:2019kee, SpECwebsite, Scheel:2008rj, Szilagyi:2009qz,
  Lovelace:2010ne, Buchman:2012dw, Hemberger:2012jz, Scheel:2014ina,
  Szilagyi:2014fna}, as well as other applications in relativistic
astrophysics~\cite{Bonazzola:1998ge, Meringolo:2021yjh, Hilditch:2015aba,
  Rashti:2021ihv, Meringolo:2020jsu, Tichy:2009zr}. The Spectral Einstein Code
(\texttt{SpEC})~\cite{SpECwebsite} performs binary neutron star merger
simulations by solving the spacetime using pseudospectral methods and the
magnetohydrodynamics (MHD) using finite difference methods~\cite{Duez:2007bz,
  Duez:2008rb, Duez:2009yy, Foucart:2010eq, Muhlberger:2014pja, Tacik:2015tja,
  Foucart:2015gaa, Haas:2016cop}. However, these use completely separate grids
requiring interpolation between them at every time/sub step. This interpolation
adds non-trivial cost, though more importantly, \texttt{SpEC}'s use of large
spectral elements causes significant load-imbalance and prohibitive cost as
resolution is increased. This is because the spacetime is only two derivatives
smoother than the MHD solution, and so the spectral approximation is less
accurate at the stellar surfaces causing \texttt{SpEC}'s adaptive mesh
refinement algorithm\cite{Szilagyi:2014fna} to significantly increase the number
of grid points in these regions. SpEC has leveraged its use of pseudospectral
methods to produce relatively low-cost 10-15 orbits long simulations of binary
neutron star (BNS) and black hole-neutron star
(BHNS)~\cite{Foucart:2018lhe,Foucart:2020xkt,Knight:2023kqw} mergers with
accuracy comparable to state-of-the art finite difference
codes~\cite{Radice:2015nva,Kiuchi:2017pte,
  Dietrich:2018phi,Gonzalez:2022mgo}. Given its scaling issues when attempting
higher resolution simulations, however, producing the longer, higher accuracy
BNS and BHNS waveforms needed by next-generation gravitational wave
detectors~\cite{PhysRevResearch.2.023151} with SpEC would be impractical.  The
same issues arise when attempting to capture the growth of magnetic fields due
to MHD instabilities during and after merger, as well as the expected dynamo
processes leading to the production of a large scale, organized magnetic field
from the small scale field generated by these instabilities. Recent simulations
have demonstrated the transfer of magnetic energy from small to large scales in
merger simulations~\cite{Aguilera-Miret:2023qih,Kiuchi:2023obe,Combi:2023yav},
yet even the highest resolution simulations available do not show clear
convergence of the magnetic field evolution during and after
merger~\cite{Kiuchi:2015sga,Palenzuela:2022kqk}. The need to perform
high-resolution MHD simulations is particularly acute given the importance of
the large scale structure of the magnetic field on matter ejection and the
electromagnetic signals powered by neutron star
mergers~\cite{Christie:2019lim,deHaas:2022ytm}.

By using the same method (DG or finite difference) in each element for both the
spacetime and MHD, \texttt{SpECTRE} is able to achieve robust convergence during
the inspiral phase while avoiding the scaling issues limiting \texttt{SpEC}'s
ability to perform high-resolution simulations. Recently,
Ref.~\cite{Dumbser:2023see} successfully performed long-term simulations of
static spacetimes using a DG-finite volume hybrid method to evolve the
Einstein-Euler system. However, inspiral and merger simulations were not
presented for black hole or neutron star binaries. There are several other
next-generation numerical relativity codes using DG methods, including
\texttt{bamps}\cite{Hilditch:2015aba, Renkhoff:2023nfw} and
\texttt{Nmesh}\cite{Tichy:2022hpa}. Other next-generation codes like
\texttt{GR-Athena++}\cite{Cook:2023bag, Daszuta:2024chz},
\texttt{Athena-K}\cite{AthenaK, 2024arXiv240916053S, Zhu:2024utz,
  Fields:2024pob},
\texttt{Parthenon}\cite{Parthenon}, \texttt{GR-Chombo}\cite{Clough:2015sqa},
\texttt{Carpet-X}\cite{Schnetter:2024},
\texttt{AsterX}\cite{2023APS..APRF08008K},
\texttt{GRAM-X}\cite{Shankar:2022ful}, and \texttt{Dendro-GR}\cite{10046060,
  Fernando:2018mov, Fernando:2022php} use FD methods (\texttt{Dendro-GR}
combines this with wavelet-based adaptive mesh refinement) but are aimed at
using Graphics Processing Units (GPUs). The code \texttt{SPHINCS\
  BSSN}\cite{Diener:2022hui} uses smooth particle hydrodynamics to evolve the
relativistic hydrodymanics and FD methods for the spacetime.

In this paper we present the improvements necessary for \texttt{SpECTRE} to
simulate the inspiral and merger of a binary neutron star system using DG
methods. We will present results of collapsing simulations in future work since
several improvements to our algorithm currently in development will be
necessary. In \S\ref{sec:equations} we briefly review the generalized harmonic
(GH) equations. In \S\ref{sec:dgscl DG-FD hybrid} we provide details about the
improvements necessary since~\cite{Deppe:2021ada} to successfully simulate the
inspiral and merger of two neutron stars. In \S\ref{sec:dgscl numerical results}
we present numerical results from simulations of a TOV star, rotating neutron
star, and a binary neutron star merger, including the first gravitational
waveforms extracted using Cauchy-Characteristic
Evolution~\cite{2020PhRvD.102d4052M, Moxon:2021gbv, 1996PhRvD..54.6153B,
  1999bhgr.conf..383B, 2005CQGra..22.2393B, 2015CQGra..32b5008H,
  2015CQGra..32w5018H, 2016LRR....19....2B, 2016CQGra..33p5001C,
  2016CQGra..33v5007H, 2020PhRvD.102b4004B}. All simulations are done using the
open-source code \texttt{SpECTRE}~\cite{Kidder:2016hev, Deppe:2021ada,
  spectrecode_2024} using the scheme presented here. We conclude in
\S\ref{sec:dgscl conclusions}.

\section{Equations of Motion}
\label{sec:equations}

We adopt the standard 3+1 form of the spacetime metric,~(see,
e.g.,~\cite{Baumgarte:2010ndz, 2013rehy.book.....R}),
\begin{eqnarray}
  \label{eq:spacetime metric}
  ds^2 &= g_{ab}dx^a dx^b =-\alpha^2 dt^2 + \gamma_{ij}
         \left(dx^i+\beta^i dt\right) \left(dx^j +\beta^j dt\right),
\end{eqnarray}
where $\alpha$ is the lapse, $\beta^i$ the shift vector, and $\gamma_{ij}$ is
the spatial metric.  We use the Einstein summation convention, summing over
repeated indices.  Latin indices from the first part of the alphabet
$a,b,c,\ldots$ denote spacetime indices ranging from $0$ to $3$, while Latin
indices $i,j,\ldots$ are purely spatial, ranging from $1$ to $3$. We work in
units where $c = G = M_{\odot} = 1$, and use geometrized Heaviside-Lorentz units
where the magnetic field is rescaled by $1/\sqrt{4\pi}$ compared to Gaussian
units.

We refer the reader to the literature~\cite{2006ApJ...637..296A, Font:2008fka,
  Baumgarte:2010ndz} for a detailed description of the equations of general
relativistic magnetohydrodynamics~(GRMHD) and their implementation in
\texttt{SpECTRE}. The generalized harmonic (GH) equations are given
by\cite{Lindblom:2005qh},
\begin{eqnarray}
  \label{eq:fosh gh metric evolution}
  \partial_t g_{ab}
  &=\left(1+\gamma_1\right)\beta^k\partial_k g_{ab}
    -\alpha \Pi_{ab}-\gamma_1\beta^i\Phi_{iab}, \\
  \label{eq:fosh gh metric derivative evolution}
  \partial_t\Phi_{iab}
  &=\beta^k\partial_k\Phi_{iab} - \alpha \partial_i\Pi_{ab}
    + \alpha \gamma_2\partial_ig_{ab}
    +\frac{1}{2}\alpha n^c n^d\Phi_{icd}\Pi_{ab} \nonumber \\
  &+ \alpha \gamma^{jk}n^c\Phi_{ijc}\Phi_{kab}
    -\alpha \gamma_2\Phi_{iab},\\
  \label{eq:fosh gh metric conjugate evolution}
  \partial_t\Pi_{ab}
  &=\beta^k\partial_k\Pi_{ab} - \alpha \gamma^{ki}\partial_k\Phi_{iab}
    + \gamma_1\gamma_2\beta^k\partial_kg_{ab} \nonumber \\
  &+2\alpha g^{cd}\left(\gamma^{ij}\Phi_{ica}\Phi_{jdb}
    - \Pi_{ca}\Pi_{db} - g^{ef}\Gamma_{ace}\Gamma_{bdf}\right) \nonumber \\
  &-2\alpha \nabla_{(a}H_{b)}
    - \frac{1}{2}\alpha n^c n^d\Pi_{cd}\Pi_{ab}
    - \alpha n^c \Pi_{ci}\gamma^{ij}\Phi_{jab} \nonumber \\
  &+\alpha \gamma_0\left(2\delta^c{}_{(a} n_{b)}
    - g_{ab}n^c\right)\mathcal{C}_c -\gamma_1\gamma_2 \beta^i\Phi_{iab}
    \nonumber \\
  &-16\pi \alpha \left(T_{ab} - \frac{1}{2}g_{ab}T^c{}_c\right),
\end{eqnarray}
where $g_{ab}$ is the spacetime metric, $\Phi_{iab}=\partial_i g_{ab}$,
$\Pi_{ab} = n^c\partial_cg_{ab}$, $n^a$ is the unit normal vector to the spatial
slice, $\gamma_0$ damps the 1-index or gauge constraint
$\mathcal{C}_a=H_a+\Gamma_a$, $\gamma_1$ controls the linear degeneracy of the
system, $\gamma_2$ damps the 3-index constraint
$\mathcal{C}_{iab}=\partial_i g_{ab}-\Phi_{iab}$, $\Gamma_{abc}$ are the
spacetime Christoffel symbols of the first kind, $\Gamma_a=g^{bc}\Gamma_{bca}$,
and $T_{ab}$ is the stress-energy tensor.  The gauge source function $H_a$ can
be any arbitrary function depending only upon the spacetime coordinates $x^a$
and $g_{ab}$ (but not derivatives of $g_{ab}$).  For the GRMHD system the
trace-reversed stress-energy tensor that sources $\Pi_{ab}$ is given by
\begin{eqnarray}
  \label{eq:fosh grmhd trace reversed}
  T_{a b} - \frac{1}{2} g_{a b} T^c{}_c
  = \left(\rho h + b^2\right) u_a u_b
  + \left[\frac{1}{2}\left(\rho h + b^2\right) - p\right] g_{a b}
  - b_a b_b,
\end{eqnarray}
where $u_a$ is the four-velocity of the fluid, $\rho$ is the baryon rest mass
density, $p$ the fluid pressure, $h$ the specific enthalpy, and
$b^a=-\frac{1}{2}\epsilon^{abcd}F_{cd}u_b$ with
$\epsilon_{abcd}=\sqrt{-g}[abcd]$, $g$ the determinant of the spacetime metric
and $[abcd]=\pm1$ with $[0123]=+1$ is the flat-space antisymmetric symbol.

\section{Discontinuous Galerkin-finite difference hybrid
  method in dynamical spacetimes\label{sec:dgscl DG-FD hybrid}}
In this section we present our DG-FD hybrid method improvements necessary to
simulate dynamical spacetimes. The reader is referred to~\cite{Deppe:2021ada}
for the original algorithm and to~\cite{Kim:2024mau} for improvements developed
for simulating general relativistic force-free electrodynamics.

\subsection{Generalized harmonic spectral filter\label{sec:gh spectral filter}}
We use an exponential filter applied to the spectral coefficient $c_i$ in order
to reduce and eliminate aliasing-driven instabilities. Specifically, for a
1d spectral expansion
\begin{eqnarray}
  \label{eq:u spectral expansion}
  u(x)=\sum_{i=0}^{N}c_i P_i(x),
\end{eqnarray}
where $P_i(x)$ are the Legendre polynomials, we use the filter
\begin{eqnarray}
  \label{eq:exponential filter}
  c_i \to c_i \exp\left[-a\left(\frac{i}{N}\right)^{2b}\right].
\end{eqnarray}
We choose the parameters $a=36$ and $b=64$ so that only the highest spectral
mode is filtered. We only apply the filter to the GH variables $g_{ab}$,
$\Phi_{iab}$ and $\Pi_{ab}$. Note that the filter drops the order of convergence
for the GH variables from $\mathcal{O}(N+1)$ to $\mathcal{O}(N)$ on the DG grid,
but is necessary for stability.

\subsection{Generalized harmonic finite difference method\label{sec:fd gh}}
When FD is necessary, we discretize the GH system using standard cell-centered
central FD methods\footnote{See, e.g.~\cite{NumericalRecipes, BaumgarteShapiro,
    RezzollaBook} for a pedagogical overview.}. In general, the order of
accuracy of the FD derivatives for the GH system is two orders higher than that
of the GRMHD system. That is, if we use second-order monotonized central
reconstruction~\cite{VanLeer:1977}, we use fourth-order FD derivatives of the GH
system. We apply Kreiss-Oliger dissipation~\cite{Kreiss:1973} as a filter to the
GH variables before taking numerical derivatives and evaluating the time
derivatives. Specifically, the filtered variable $\tilde{u}$ is given by
\begin{eqnarray}
  \label{eq:fd dissipation filter}
  \tilde{u}_{\hat{\imath}} = u_{\hat{\imath}} + \epsilon
  F^{(m)}(u_{\hat{\imath}}),
\end{eqnarray}
for Kreiss-Oliger operator $F^{(m)}$ and $\epsilon\in[0,1]$. The
subscript $\hat{\imath}$ refers to a grid point index. We use the filter
$F^{(5)}$ when using fourth-order FD derivatives where $F^{(5)}$ is given by
\begin{eqnarray}
  \label{eq:fd dissipation filter 5}
  F^{(5)} u_{\hat{\imath}}= -\frac{375}{8} \left(
  \frac{1}{336} (u_{\hat{\imath}-2} + u_{\hat{\imath}+2}) -
  \frac{1}{84} (u_{y-1} + u_{\hat{\imath}+1}) +
  \frac{1}{56} u_{\hat{\imath}}\right).
\end{eqnarray}
The advantage of this approach is that the number of ghost zones
\blue{(i.e., the number of grid points that are sent by neighboring elements to
compute derivatives and perform reconstruction)} is not
increased for dissipation, and the GH system is still solved at a higher order
than the GRMHD system.

We reconstruct the metric variables to subcell interfaces at the same order as
the hydro variables. That is, if we use the fifth-order positivity-preserving
adaptive-order (PPAO) scheme\cite{Deppe:2023qxa} to reconstruct the GRMHD
variables, then we use unlimited fifth-order reconstruction for the metric.

\subsection{GRMHD finite difference method\label{sec:fd grmhd}}

The overall method is very similar to that presented in~\cite{Deppe:2021ada,
  Deppe:2023qxa}. However, instead of reconstructing the pressure $p$ we now
reconstruct the ``temperature'' $T$ of the fluid. This is because the
temperature is an independent variable in equation of state tables, so even if
there are numerical errors, as long as the temperature remains positive the
reconstructed primitive state is ``reasonable''.

\subsection{Curved mesh finite difference method\label{sec:fd curved mesh}}

\texttt{SpECTRE} solves hyperbolic systems of equations of the form
\begin{eqnarray}
  \label{eq:model pde}
  \partial_t u + \partial_i F^i(u) + B^i(u)\partial_i u = S(u),
\end{eqnarray}
where $u$ are the evolved variables, $F^i(u)$ the fluxes, $B^i(u)$
non-conservative products, and $S(u)$ source terms. In our DG-FD method the
computational domain is divided up into non-overlapping elements or cells, which
we denote by $\Omega_k$. This allows us to write the system~\eref{eq:model pde}
in semi-discrete form where time remains continuous. In the DG method one
integrates the evolution equations~\eref{eq:model pde} against
spatial basis functions of degree $N$, which we denote by
$\phi_{\breve{\imath}}$. We index the basis functions and collocation points of
the DG scheme with breve Latin indices, e.g.~$\breve{\imath}, \breve{\jmath},
\breve{k}$. The basis functions are defined in the reference coordinates of each
element, which we denote by $\xi^{\hat{\imath}}\in\{\xi, \eta, \zeta\}$.  We use
hatted indices to denote tensor components in the reference frame. The reference
coordinates are mapped to the physical coordinates using the general function
\begin{eqnarray}
  \label{eq:dgscl time independent coordinate map}
  x^i=x^i(\xi^{\hat{\imath}}, t).
\end{eqnarray}
Since the reference coordinates $\xi^{\hat{\imath}}$ are Cartesian, applying a
FD scheme is comparatively straightforward to implement in the reference
coordinates rather than the physical or inertial coordinates that the hyperbolic
equations are written in.

In order to support FD on curved meshes \texttt{SpECTRE} now solves the
equations in the form
\begin{eqnarray}
  \label{eq:subcell logical frame conservation law}
  \frac{\partial u}{\partial t}
  + \frac{1}{J}\partial_{\hat{\imath}}
  \left[
  J\frac{\partial \xi^{\hat{\imath}}}{\partial x^i}
  \left(F^i-u v^i_g\right)\right]
  = S - u\partial_i v^i_g,
\end{eqnarray}
where $v^i_g$ is the grid or mesh velocity and we can identify
$F^{\hat{\imath}}=J\frac{\partial \xi^{\hat{\imath}}}{\partial x^i} F^i$ as the
``locally Cartesian flux'' in the reference coordinates. This is analogous to
how DG schemes are formulated~\cite{Teukolsky:2015ega, Deppe:2021ada}. In
practice we compute the mesh velocity divergence on the DG grid and project it
to the FD grid. While this form is somewhat different from the strong form used
by our DG solver, we can still rewrite the equations in a form that naturally
hybridizes with a DG solver. In particular, the boundary corrections $G$ in a DG
scheme are essentially $n_iF^i$ where $n_i$ is the normal covector to the
spatial element interface and in the logical $\hat{\imath}$ direction is given
by
\begin{eqnarray}
  \label{eq:normal covector}
  n^{(\hat{\imath})}_i=\frac{\partial \xi^{(\hat{\imath})}}{\partial x^i}
  \frac{1}{\sqrt{\frac{\partial \xi^{(\hat{\imath})}}{\partial x^j}
  \gamma^{jk}\frac{\partial \xi^{(\hat{\imath})}}{\partial x^k}}}
  =J\frac{\partial \xi^{(\hat{\imath})}}{\partial x^i}
  \frac{1}{\sqrt{J\frac{\partial \xi^{(\hat{\imath})}}{\partial x^j}
  \gamma^{jk}J\frac{\partial \xi^{(\hat{\imath})}}{\partial x^k}}}.
\end{eqnarray}
With $G^{(\hat{\imath})}$ as the boundary correction or numerical flux at the
interface in direction $\hat{\imath}$, possibly including a high-order
correction\cite{DelZanna:2007pk, Most:2019kfe}, we can write the discretized FD
evolution equation as
\begin{eqnarray}
  \label{eq:subcell logical frame difference equation}
  \frac{\partial u_{(\xi,\eta,\zeta)}}{\partial t} &= S_{(\xi,\eta,\zeta)}
  -u_{(\xi,\eta,\zeta)} \mathcal{P}(\partial_i v^i_g)_{(\xi,\eta,\zeta)}
  \nonumber \\
  &- \frac{1}{J_{(\xi,\eta,\zeta)}}\left[
    \frac{\left(|n^{(\hat{\xi})}|G^{\hat{(\xi)}}\right)_{(\xi +
    1/2,\eta,\zeta)}
    -\left(|n^{(\hat{\xi})}|G^{\hat{(\xi)}}\right)_{(\xi - 1/2,\eta,\zeta)}
    }{\Delta \xi}\right.\nonumber\\
  &\left.+
    \frac{\left(|n^{(\hat{\eta})}|G^{\hat{(\eta)}}\right)_{(\xi,\eta +
    1/2,\zeta)}
    -\left(|n^{(\hat{\eta})}|G^{\hat{(\eta)}}\right)_{(\xi,\eta - 1/2,\zeta)}
    }{\Delta \eta}
    \right.\nonumber\\
  &\left.+
    \frac{\left(|n^{(\hat{\zeta})}|G^{\hat{(\zeta)}}\right)_{(\xi,\eta,\zeta +
    1/2)}
    -\left(|n^{(\hat{\zeta})}|G^{\hat{(\zeta)}}\right)_{(\xi,\eta,\zeta -
    1/2)}
    }{\Delta \zeta}
    \right],
\end{eqnarray}
where $\mathcal{P}$ is the projection operator from the DG to the FD grid as
defined in~\cite{Deppe:2021ada} and
$|n^{(\hat{\imath})}|=\sqrt{J\frac{\partial \xi^{(\hat{\imath})}}{\partial
    x^j}\gamma^{jk}J\frac{\partial \xi^{(\hat{\imath})}}{\partial x^k}}$ or
$|n^{(\hat{\imath})}|=\sqrt{\frac{\partial \xi^{(\hat{\imath})}}{\partial
    x^j}\gamma^{jk}\frac{\partial \xi^{(\hat{\imath})}}{\partial x^k}}$ depending
on which normal vector form is chosen. Since the correction $G^{(\hat{\imath})}$
is exactly what is used in a DG scheme we can straightforwardly hybridize the
two schemes in a conservative manner, independent of the exact DG or FD
formulation used. The primary reason for using the discretized form in
Eq.~\ref{eq:subcell logical frame difference equation} is for easier
implementation on curved meshes since in this form the equations are simply the
standard Cartesian evolution equations.

A non-trivial challenge on curved meshes is populating ghost zones. In
\texttt{SpECTRE} we divide the computational domain into a set of conforming
collections of elements called \textit{blocks}. Each block has Cartesian
block-logical coordinates $[-1,1]^3$. These coordinates are then mapped by a
possibly time-dependent map to the inertial frame in which the evolution
equations are written. This is a standard approach in the spectral finite
element method community and is similar to what \texttt{SpEC}
uses\cite{Scheel:2006gg, Hemberger:2012jz}. An example domain is a 2d disk made
up of five deformed rectangles. One square is in the middle surrounded by 4
wedges. The specific challenge is reconstruction at block boundaries since the
block-logical coordinates of two blocks in general do not align. This requires
some form of interpolation to populate ghost points for the neighbor. Currently
we use simple trilinear interpolation. However, we are developing an approach
based on high-order limited interpolation inspired by~\cite{CHEN2016604}.

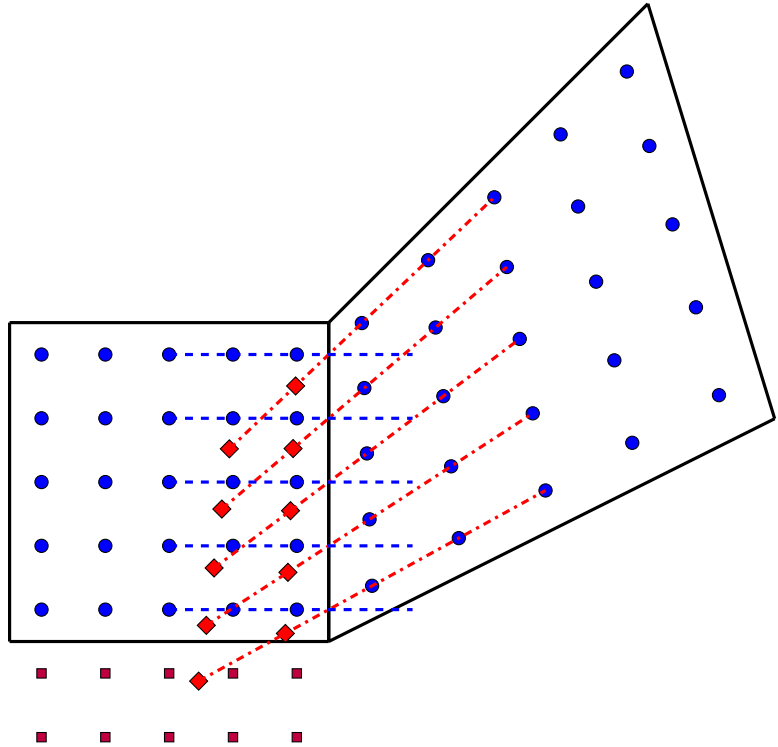
\begin{figure}[h]
  \centering

  \begin{tikzpicture}[ hatch distance/.store in=\hatchdistance, hatch
      distance=30pt, scale=12, square/.style={regular polygon,regular polygon sides=4}]

      \node[circle,draw=black, fill=blue, inner sep=0pt,minimum size=5pt] (b) at (0.388909, 0.388909) {};
      \node[circle,draw=black, fill=blue, inner sep=0pt,minimum size=5pt] (b) at (0.459619, 0.388909) {};
      \node[circle,draw=black, fill=blue, inner sep=0pt,minimum size=5pt] (b) at (0.53033, 0.388909) {};
      \node[circle,draw=black, fill=blue, inner sep=0pt,minimum size=5pt] (b) at (0.601041, 0.388909) {};
      \node[circle,draw=black, fill=blue, inner sep=0pt,minimum size=5pt] (b) at (0.671751, 0.388909) {};
      \node[circle,draw=black, fill=blue, inner sep=0pt,minimum size=5pt] (b) at (0.388909, 0.459619) {};
      \node[circle,draw=black, fill=blue, inner sep=0pt,minimum size=5pt] (b) at (0.459619, 0.459619) {};
      \node[circle,draw=black, fill=blue, inner sep=0pt,minimum size=5pt] (b) at (0.53033, 0.459619) {};
      \node[circle,draw=black, fill=blue, inner sep=0pt,minimum size=5pt] (b) at (0.601041, 0.459619) {};
      \node[circle,draw=black, fill=blue, inner sep=0pt,minimum size=5pt] (b) at (0.671751, 0.459619) {};
      \node[circle,draw=black, fill=blue, inner sep=0pt,minimum size=5pt] (b) at (0.388909, 0.53033) {};
      \node[circle,draw=black, fill=blue, inner sep=0pt,minimum size=5pt] (b) at (0.459619, 0.53033) {};
      \node[circle,draw=black, fill=blue, inner sep=0pt,minimum size=5pt] (b) at (0.53033, 0.53033) {};
      \node[circle,draw=black, fill=blue, inner sep=0pt,minimum size=5pt] (b) at (0.601041, 0.53033) {};
      \node[circle,draw=black, fill=blue, inner sep=0pt,minimum size=5pt] (b) at (0.671751, 0.53033) {};
      \node[circle,draw=black, fill=blue, inner sep=0pt,minimum size=5pt] (b) at (0.388909, 0.601041) {};
      \node[circle,draw=black, fill=blue, inner sep=0pt,minimum size=5pt] (b) at (0.459619, 0.601041) {};
      \node[circle,draw=black, fill=blue, inner sep=0pt,minimum size=5pt] (b) at (0.53033, 0.601041) {};
      \node[circle,draw=black, fill=blue, inner sep=0pt,minimum size=5pt] (b) at (0.601041, 0.601041) {};
      \node[circle,draw=black, fill=blue, inner sep=0pt,minimum size=5pt] (b) at (0.671751, 0.601041) {};
      \node[circle,draw=black, fill=blue, inner sep=0pt,minimum size=5pt] (b) at (0.388909, 0.671751) {};
      \node[circle,draw=black, fill=blue, inner sep=0pt,minimum size=5pt] (b) at (0.459619, 0.671751) {};
      \node[circle,draw=black, fill=blue, inner sep=0pt,minimum size=5pt] (b) at (0.53033, 0.671751) {};
      \node[circle,draw=black, fill=blue, inner sep=0pt,minimum size=5pt] (b) at (0.601041, 0.671751) {};
      \node[circle,draw=black, fill=blue, inner sep=0pt,minimum size=5pt] (b) at (0.671751, 0.671751) {};

      \draw[-, very thick] (0.353553,0.353553) -- (0.353553,0.707107);
      \draw[-, very thick] (0.353553,0.707107) -- (0.707107,0.707107);
      \draw[-, very thick] (0.707107,0.707107) -- (0.707107,0.353553);
      \draw[-, very thick] (0.707107,0.353553) -- (0.353553,0.353553);

      \node[circle,draw=black, fill=blue, inner sep=0pt,minimum size=5pt] (b) at (0.755145, 0.41533) {};
      \node[circle,draw=black, fill=blue, inner sep=0pt,minimum size=5pt] (b) at (0.851222, 0.468172) {};
      \node[circle,draw=black, fill=blue, inner sep=0pt,minimum size=5pt] (b) at (0.947299, 0.521015) {};
      \node[circle,draw=black, fill=blue, inner sep=0pt,minimum size=5pt] (b) at (1.04338, 0.573857) {};
      \node[circle,draw=black, fill=blue, inner sep=0pt,minimum size=5pt] (b) at (1.13945, 0.626699) {};
      \node[circle,draw=black, fill=blue, inner sep=0pt,minimum size=5pt] (b) at (0.752312, 0.489003) {};
      \node[circle,draw=black, fill=blue, inner sep=0pt,minimum size=5pt] (b) at (0.842724, 0.54777) {};
      \node[circle,draw=black, fill=blue, inner sep=0pt,minimum size=5pt] (b) at (0.933135, 0.606538) {};
      \node[circle,draw=black, fill=blue, inner sep=0pt,minimum size=5pt] (b) at (1.02355, 0.665305) {};
      \node[circle,draw=black, fill=blue, inner sep=0pt,minimum size=5pt] (b) at (1.11396, 0.724072) {};
      \node[circle,draw=black, fill=blue, inner sep=0pt,minimum size=5pt] (b) at (0.749429, 0.562072) {};
      \node[circle,draw=black, fill=blue, inner sep=0pt,minimum size=5pt] (b) at (0.834074, 0.625555) {};
      \node[circle,draw=black, fill=blue, inner sep=0pt,minimum size=5pt] (b) at (0.918718, 0.689039) {};
      \node[circle,draw=black, fill=blue, inner sep=0pt,minimum size=5pt] (b) at (1.00336, 0.752522) {};
      \node[circle,draw=black, fill=blue, inner sep=0pt,minimum size=5pt] (b) at (1.08801, 0.816006) {};
      \node[circle,draw=black, fill=blue, inner sep=0pt,minimum size=5pt] (b) at (0.746575, 0.634588) {};
      \node[circle,draw=black, fill=blue, inner sep=0pt,minimum size=5pt] (b) at (0.82551, 0.701684) {};
      \node[circle,draw=black, fill=blue, inner sep=0pt,minimum size=5pt] (b) at (0.904446, 0.768779) {};
      \node[circle,draw=black, fill=blue, inner sep=0pt,minimum size=5pt] (b) at (0.983381, 0.835874) {};
      \node[circle,draw=black, fill=blue, inner sep=0pt,minimum size=5pt] (b) at (1.06232, 0.902969) {};
      \node[circle,draw=black, fill=blue, inner sep=0pt,minimum size=5pt] (b) at (0.743804, 0.706614) {};
      \node[circle,draw=black, fill=blue, inner sep=0pt,minimum size=5pt] (b) at (0.817199, 0.776339) {};
      \node[circle,draw=black, fill=blue, inner sep=0pt,minimum size=5pt] (b) at (0.890593, 0.846064) {};
      \node[circle,draw=black, fill=blue, inner sep=0pt,minimum size=5pt] (b) at (0.963988, 0.915788) {};
      \node[circle,draw=black, fill=blue, inner sep=0pt,minimum size=5pt] (b) at (1.03738, 0.985513) {};

      \draw[-, very thick] (0.707107,0.353553) -- (0.707107,0.707107);
      \draw[-, very thick] (0.707107,0.707107) -- (1.06066,1.06066);
      \draw[-, very thick] (1.06066,1.06066) -- (1.20115,0.600575);
      \draw[-, very thick] (1.20115,0.600575) -- (0.707107,0.353553);

      \node[diamond,draw=black, fill=red, inner sep=0pt,minimum size=7pt] (b) at (0.659068, 0.362488) {};
      \node[diamond,draw=black, fill=red, inner sep=0pt,minimum size=7pt] (b) at (0.562991, 0.309645) {};

      \node[diamond,draw=black, fill=red, inner sep=0pt,minimum size=7pt] (b) at (0.661901, 0.430236) {};
      \node[diamond,draw=black, fill=red, inner sep=0pt,minimum size=7pt] (b) at (0.57149, 0.371468) {};

      \node[diamond,draw=black, fill=red, inner sep=0pt,minimum size=7pt] (b) at (0.664784, 0.498588) {};
      \node[diamond,draw=black, fill=red, inner sep=0pt,minimum size=7pt] (b) at (0.58014, 0.435105) {};

      \node[diamond,draw=black, fill=red, inner sep=0pt,minimum size=7pt] (b) at (0.667639, 0.567493) {};
      \node[diamond,draw=black, fill=red, inner sep=0pt,minimum size=7pt] (b) at (0.588703, 0.500398) {};

      \node[diamond,draw=black, fill=red, inner sep=0pt,minimum size=7pt] (b) at (0.670409, 0.636889) {};
      \node[diamond,draw=black, fill=red, inner sep=0pt,minimum size=7pt] (b) at (0.597015, 0.567164) {};

      \draw[very thick, dash dot, draw=red] (0.562991, 0.309645) -- (0.947299, 0.521015);
      \draw[very thick, dash dot, draw=red] (0.57149, 0.371468) -- (0.933135, 0.606538);
      \draw[very thick, dash dot, draw=red] (0.58014, 0.435105) -- (0.918718, 0.689039);
      \draw[very thick, dash dot, draw=red] (0.588703, 0.500398) -- (0.904446, 0.768779);
      \draw[very thick, dash dot, draw=red] (0.597015, 0.567164) -- (0.890593, 0.846064);

      \draw[very thick, dashed, draw=blue] (0.53033, 0.388909) -- (0.8, 0.388909);
      \draw[very thick, dashed, draw=blue] (0.53033, 0.459619) -- (0.8, 0.459619);
      \draw[very thick, dashed, draw=blue] (0.53033, 0.53033) -- (0.8, 0.53033);
      \draw[very thick, dashed, draw=blue] (0.53033, 0.601041) -- (0.8, 0.601041);
      \draw[very thick, dashed, draw=blue] (0.53033, 0.671751) -- (0.8, 0.671751);

      \node[square,draw=black, fill=purple, inner sep=0pt,minimum size=5pt] (b) at (0.388909, 0.247487) {};
      \node[square,draw=black, fill=purple, inner sep=0pt,minimum size=5pt] (b) at (0.459619, 0.247487) {};
      \node[square,draw=black, fill=purple, inner sep=0pt,minimum size=5pt] (b) at (0.53033, 0.247487) {};
      \node[square,draw=black, fill=purple, inner sep=0pt,minimum size=5pt] (b) at (0.601041, 0.247487) {};
      \node[square,draw=black, fill=purple, inner sep=0pt,minimum size=5pt] (b) at (0.671751, 0.247487) {};
      \node[square,draw=black, fill=purple, inner sep=0pt,minimum size=5pt] (b) at (0.388909, 0.318198) {};
      \node[square,draw=black, fill=purple, inner sep=0pt,minimum size=5pt] (b) at (0.459619, 0.318198) {};
      \node[square,draw=black, fill=purple, inner sep=0pt,minimum size=5pt] (b) at (0.53033, 0.318198) {};
      \node[square,draw=black, fill=purple, inner sep=0pt,minimum size=5pt] (b) at (0.601041, 0.318198) {};
      \node[square,draw=black, fill=purple, inner sep=0pt,minimum size=5pt] (b) at (0.671751, 0.318198) {};

    \end{tikzpicture}
    \caption{An illustration of the ghost points needed for the FD scheme where
      neighboring elements do not have aligned coordinate axes in their
      reference frames. Blue circles denote the cell-centered FD points in the
      two elements whose ghost cells are being exchanged, purple squares denote
      the cell-centered FD points in a neighboring element not used for ghost
      cell exchange in our current algorithm, and diamonds denote the ghost
      cells needed for reconstruction in the element on the right. The diagonal
      dash-dotted lines trace out lines of constant reference coordinates in the
      element on the right and dashed lines in the element on the left. Notice
      that the dashed and dash-dotted lines intersect on the element
      boundary.\label{fig:curved mesh illustration}}
\end{figure}

In figure~\ref{fig:curved mesh illustration} we show an example of a 2d domain
where the logical coordinates are not aligned. The left element must interpolate
its solution to the red diamonds when populating the ghost zones for the right
element. We currently use compact linear interpolation where neighbor points are
not used during the interpolation. Not using neighbor points means that if the
ghost zones lie outside the region enclosed by the grid points, extrapolation is
used.

\subsection{Troubled-cell indicators\label{sec:dgscl TCI}}
One of the most important parts of the DG-FD hybrid method is the
\blue{troubled-cell indicator (TCI)} that determines when to switch from DG to
FD and back. We still use the relaxed discrete maximum principle (RDMP) as
discussed in~\cite{Dumbser2014a, Deppe:2021ada}. Specifically, the RDMP requires
that
\begin{eqnarray}
  \label{eq:dgscl rdmp}
  \min_{\mathcal{N}}\left[u(t^n)\right]
  - \delta
  \le
  u^\star(t^{n+1})
  \le
  \max_{\mathcal{N}}\left[u(t^n)\right] + \delta,
\end{eqnarray}
where $\mathcal{N}$ are either the Neumann or Voronoi neighbors plus the element
itself, $\delta$ is a parameter defined below that relaxes the discrete maximum
principle, $u$ are the conserved variables, and $u^{\star}(t^{n+1})$ is a
candidate solution at time $t^{n+1}$ computed using an unlimited DG scheme. When
computing $\max(u)$ and $\min(u)$ over an element using DG, we first project the
DG solution to the FD grid and then compute the maximum and minimum over
\textit{both} the DG and FD grid. However, when an element is using FD we
compute the maximum and minimum over the FD grid only. The maximum and minimum
values of $u^\star$ are computed in the same manner as those of $u$. The
parameter $\delta$ used to relax the discrete maximum principle is given by:
\begin{eqnarray}
  \label{eq:dgscl rdmp delta}
  \delta =
  \max\left(\delta_{0},\epsilon
  \left\{\max_{\mathcal{N}}\left[u(t^n)\right]
  - \min_{\mathcal{N}}\left[u(t^n)\right]\right\}\right),
\end{eqnarray}
where, as in~\cite{Dumbser2014a}, we take $\delta_{0}=10^{-7}$ and
$\epsilon=10^{-3}$. If the condition~\ref{eq:dgscl rdmp} is satisfied, we say
the variable $u$ passes the RDMP.

We also continue to use the Persson indicator~\cite{PerssonTci}; however, we
have changed the details. Specifically, consider a variable $u$ with a 1d
spectral decomposition:
\begin{eqnarray}
  \label{eq:dgscl Persson U expansion}
  u(x)=\sum_{{\hat{\imath}}=0}^{N}c_{\hat{\imath}} P_{\hat{\imath}}(x),
\end{eqnarray}
where in our case $P_{\hat{\imath}}(x)$ are Legendre polynomials, and
$c_{\hat{\imath}}$ are the spectral coefficients. The Persson TCI essentially
monitors the percentage of power in the highest spectral coefficient(s). To do
this, we define $\hat{u}$ as
\begin{eqnarray}
  \label{eq:dgscl Persson U hat}
  \hat{u}(x)=c_N P_N(x).
\end{eqnarray}
and check that
\begin{eqnarray}
  \label{eq:dgscl Persson indicator new}
  (N+1)^\alpha \sqrt{\sum_{{\hat{\imath}}=0}^N \hat{u}_{\hat{\imath}}^2} >
  \sqrt{\sum_{{\hat{\imath}}=0}^N
  u_{\hat{\imath}}^2},
\end{eqnarray}
where $(N+1)^\alpha$ can be precomputed and stored. We find that this
mathematically equivalent condition to our previous check~\cite{Deppe:2021ada},
\begin{eqnarray}
  \label{eq:dgscl Persson indicator old}
  s^\Omega=\log_{10}\left(\sqrt{\frac{\sum_{{\hat{\imath}}=0}^N
  \hat{u}_{\hat{\imath}}^2}{\sum_{{\hat{\imath}}=0}^N
  u_{\hat{\imath}}^2}}\right) <s^e=-\alpha_N\log_{10}(N+1),
\end{eqnarray}
is cheaper and better behaved in the limit of $u\to0$.

A significant change in handling initial data is that all elements start on the
FD grid and then evaluate the TCI to see if restriction to the DG grid is
allowed. This is particularly useful for initial data interpolated to the grid
from another grid, e.g.~when reading data from an elliptic solver such
as~\cite{Pfeiffer2003, FoucartEtAl:2008, Haas:2016, Fischer:2021voj,
  Vu:2021coj}. The TCI used on the initial FD grid is essentially identical to
the one used during the evolution described below, \S\ref{sec:dgscl FD TCI},
except for the RDMP TCI. For the RDMP TCI the candidate solution is the
restricted DG solution of the initial data.

Below we denote time levels by superscripts. For example, $u^n$ is the value of
the variable $u$ at time $t^n$ while $u^{n+1}$ is the value of the variable $u$
at time $t^{n+1}$. We also monitor several conserved magnetohydrodynamical
variables, which are defined as
\begin{eqnarray}
  \left(\begin{array}{c}
    \tilde{D} \\
    \tilde{S}_i \\
    \tilde{\tau} \\
    \tilde{B}^i
  \end{array}\right)
  = \sqrt{\gamma}
    \left(\begin{array}{c}
      \rho W \\
      (\rho h + b^2) W^2 v_i - \alpha b^0 b_i \\
      (\rho h + b^2)^* W^2 - \left[p+\frac{b^2}{2}\right] - \left(\alpha
      b^0\right)^2 - \rho W \\
      B^i
    \end{array}\right),
  \label{eq:conserved variables}
\end{eqnarray}
where $\gamma$ is the determinant of the spatial metric $\gamma_{ij}$, $v^i$ is
the spatial velocity of the fluid as measured by an observer at
rest in the spatial hypersurfaces (``Eulerian observer'') is
\begin{equation}
  \label{eq:spatial velocity}
  v^i = \frac{1}{\alpha}\left(\frac{u^i}{u^0} + \beta^i\right),
\end{equation}
with a corresponding Lorentz factor $W$
\begin{equation}
  \label{eq:Lorentz factor}
  W = - u^a n_a = \alpha u^0 = \frac{1}{\sqrt{1 - \gamma_{ij}v^i v^j}} =
  \sqrt{1+\gamma^{ij}u_i u_j},
\end{equation}
and
\begin{equation}
  \label{eq:Eulerian magnetic field}
  B^i = \dual F^{ia}n_a = \alpha \dual F^{0i}.
\end{equation}

\subsubsection{TCI on DG grid for GRMHD\label{sec:dgscl DG TCI}}

On the DG grid we require:
\begin{enumerate}
\item that $\min(\tilde{D}^{n+1})/\avg(\sqrt{\gamma^{n}})\ge D_{\min}$ on both
  the DG and the projected FD grid.
\item that $\min(\tilde{\tau}^{n+1})/\avg(\sqrt{\gamma^{n}})\ge \tau_{\min}$ on
  both the DG and the projected FD grid.
\item that $\max(\rho^{n+1})/\avg(\sqrt{\gamma^{n}})\ge \rho_{\mathrm{atm}}$
  on the DG grid. This is to ensure that we only apply the below TCI checks when
  the solution will not be reset to atmosphere since we would like to always use
  the DG solver in atmosphere.
\item that $\tilde{B}^2\le1.0 - \epsilon_B 2 \tilde{\tau}\sqrt{\gamma}$ at all
  grid points in the DG element.
\item that primitive recovery is successful.
\item if we are in atmosphere we mark the solution as admissible.
\item that $\tilde{D}$ and the pressure $p$ pass the Persson TCI.
\item that if $\max\left(\sqrt{\tilde{B}^i\delta_{ij}\tilde{B}^j}\right)$ is
  above a user-specified threshold, $\sqrt{\tilde{B}^i\delta_{ij}\tilde{B}^j}$
  satisfies the Persson TCI.
\item that the RDMP TCI passes for $\tilde{D}$, $\tilde{\tau}$, and
  $\sqrt{\tilde{B}^2}$.
\end{enumerate}
If all requirements are met, then the DG solution is admissible. We use
$\avg(\sqrt{\gamma^n})$ to reduce computational cost since we can use the same
average on both the DG and FD grid. This eliminates the need to project
$\sqrt{\gamma}$ and also reduces the amount of memory bandwidth needed.

\subsubsection{TCI on FD grid for GRMHD\label{sec:dgscl FD TCI}}
In order to switch to DG from FD, we require:
\begin{enumerate}
\item that $\min(\tilde{D}^{n+1})/\avg(\sqrt{\gamma^{n}})\ge D_{\min}$ and
  $\min(\tilde{\tau}^{n+1})/\avg(\sqrt{\gamma^{n}})\ge \tau_{\min}$ on the DG grid.
\item that we did not need to fix the conservative variables (see \S III.F
  of~\cite{Deppe:2021bhi}) if we are not in atmosphere.
\item that $\tilde{D}$ and the pressure $p$ pass the Persson TCI if we are not
  in atmosphere.
\item that the RDMP TCI passes for $\tilde{D}$, $\tilde{\tau}$, and
  $\sqrt{\tilde{B}^2}$.
\item that if $\max\left(\sqrt{\tilde{B}^i\delta_{ij}\tilde{B}^j}\right)$ is
  above a user-specified threshold, $\sqrt{\tilde{B}^i\delta_{ij}\tilde{B}^j}$
  satisfies the Persson TCI.
\end{enumerate}
If all the above checks are satisfied, then the numerical solution is
representable on the DG grid.

\subsection{Restriction from FD to DG\label{sec:dgscl dg fd restriction}}

The restriction operator $\mathcal{R}$\footnote{referred to as reconstruction
  in~\cite{Deppe:2021ada}, which we find is easily confused with the
  reconstruction done on the FD grid} that interpolates variables from the FD to
the DG grid, as presented in~\cite{Deppe:2021ada}, is a 3d operator in 3 spatial
dimensions. This means it is a matrix of size
$(N+1)^3\times(2N+1)^3$\footnote{$N$ is the degree of the DG basis and $N+1$ is
  the number of DG grid points per dimension.}, resulting in a rather expensive
matrix multiplication in the troubled-cell indicator (TCI) used on the FD grid,
where we restrict $\tilde{D}$, $p$, and optionally $\sqrt{B^i\delta_{ij}B^j}$
from the FD grid to the DG grid. This turns out to be a non-negligible expense
and so instead we apply the 1d restriction operator dimension-by-dimension. This
is a stronger constraint on the DG solution than the 3d restriction, but in
addition to the reduced cost it also guarantees that if the solution is constant
along an axis of the element on the FD grid, it will also be constant on the DG
grid. This ultimately helps reduce noise introduced through restriction.

An additional two performance improvements that reduce how frequently the TCI is
run on the FD grid were introduced after many of the simulations presented here
were already completed. These are:
\begin{enumerate}
\item When an element switches from DG to FD, enough time must elapse for the
  discontinuous feature to propagate through the troubled element before a TCI
  check is necessary. A heuristic for choosing the number of time steps to wait
  before running the TCI on the FD grid is
  \begin{equation}
    \frac{\min(\Delta x)}{\Delta t}(2 N + 1)
  \end{equation}
  where $\min(\Delta x)$ is the minimum grid spacing between FD points in the
  inertial frame, $\Delta t$ is the time step size, and $(2N+1)$ is the number
  of FD grid points per dimension in the element. For example, for a $P_5$ DG-FD
  method with $\min(\Delta x) / \Delta t\sim 2$, we should wait $\sim 22$ time
  steps before checking the TCI after switching from DG to FD.
\item Instead of checking the TCI every step after the initial check, we check
  with a specified frequency that we typically choose to be $\sim(2N-1)$ time
  steps, primarily to reduce the overhead of TCI calls. A heuristic argument for
  the frequency at which to check is not clear. Essentially, one wants to
  minimize the overhead incurred by calls to the TCI while not spending too much
  time using FD when DG would work.
\end{enumerate}
\texttt{SpECTRE} has input file options that allow controlling the two
frequencies at which the TCI is applied on the FD grid.

A third option added to the TCI, but not yet extensively tested, is requiring
the TCI to mark the solution in an element as admissible multiple times before
switching back to DG. The motivation for this is to provide additional time for
the FD solver to smooth the solution and to prevent having to switch back to the
FD grid soon after switching to DG.

All three of these methods were necessary when studying more dynamical systems
like current sheets in general relativistic force-free
electrodynamics~\cite{Kim:2024mau} and so is not just a characteristic of GRMHD,
but dynamical systems in general.

\subsection{Generalized harmonic system at DG-FD interface\label{sec:dg-fd
    interface}}

For systems in flux-conservative and flux-balanced form, stable methods for a
DG-FD hybrid scheme have been developed~\cite{Deppe:2021ada, Deppe:2021bhi,
  Kim:2024mau, DUMBSER20091731, Dumbser2014a, BOSCHERI2017449, Zanotti:2015mia,
  Fambri:2018udk, Dumbser:2023see}. These are all based on a weak form of the
system of partial differential equations. However, since the GH system is not in
flux-conservative form it is not as clear how to couple the DG and FD solver.
While weak forms for non-conservative systems exist~\cite{DALMASO1995,
  DUMBSER20091731, DUMBSER2016275, Dumbser:2017okk, 2018arXiv180803788D,
  Fambri:2018udk, Dumbser:2023see}, these formulations are not developed for FD
schemes. We opt for a simple approach. On the FD grid we use cell-centered FD
stencils using the GH variables in the ghost zones as interpolated by the DG
grid. On the DG grid we interpolate the GH variables on the FD grid to the
interface using unlimited reconstruction and then use the DG boundary correction
just as is done for flux-conservative systems. In practice we find this to be
stable except when the hybrid solver rapidly switches back and forth between the
DG and FD grids. However, we view that as an issue with the TCI and not with how
we handle the DG-FD interface for non-conservative systems. The same behavior is
observed in simulations of current sheets in general relativistic force-free
electrodynamics~\cite{Kim:2024mau} and the methods described at the end
of~\S\ref{sec:dgscl dg fd restriction} result in a robust TCI that does not
exhibit such pathological behavior.

\subsection{Outer boundary conditions\label{sec:boundary conditions}}

We impose constraint-preserving boundary conditions on the GH constraint
variables\cite{Lindblom:2005qh}, first-order
Bayliss-Turkel-type\cite{BaylissTurkel} boundary conditions on the gauge degrees
of freedom\cite{Rinne:2007ui}, and a no-incoming radiation boundary condition on
the physical degrees of freedom\cite{Lindblom:2005qh}. The boundary conditions
are imposed using the Bj{\o}rhus method\cite{Bjorhus1995, Lindblom:2005qh}. In
the future we plan to use Cauchy-Characteristic Matching to impose more
realistic boundary conditions on the incoming physical
fields\cite{Ma:2023qjn}. We impose outflow boundary conditions on the GRMHD
variables, filling ghost zones reflected about the outer boundary. For the DG
grid this means the primitive variables are simply copied from the interior
interface to the exterior one. However, we adjust the spatial
velocity. Specifically, for an outward-directed normal vector $n_i$ at the grid
points, if $n_iv^i\ge0$ we use $v^i_{\mathrm{ghost}}=v^i$ while if $n_iv^i<0$
then we set $v^i_{\mathrm{ghost}}=v^i-n^i (n_j v^j)$. These boundary conditions
allow us to stably evolve single and binary neutron star spacetimes for long
times, though our simulations are terminated before the matter reaches the outer
boundary. Because \texttt{SpECTRE}'s FD solver does not yet have the ability to
handle refinement boundaries, matter within 4 code units of the boundary of the
inner region (see \S~\ref{sec:bns mergers}) is removed from the evolution
(i.e.~the density is set to our numerical floor).

A crucial future improvement will be better handling of the matter outflows in
two key ways. First, we need to add mesh refinement support to the FD solver in
order to track matter outflows. Second, we plan to add the ability to impose
outflow boundary conditions on the matter fields inside the computational
domain. That is, rather than enforcing outflow boundary conditions in the
wavezone we impose them closer to the binary and do not evolve the GRMHD system
farther out. This is so that a larger computational domain can be used to track
the GW emission but we can ignore low-density outflows in the wavezone to reduce
computational cost.

\subsection{Constraint damping\label{sec:constraint damping}}

One non-trivial challenge in evolving the first-order GH system is choosing
constraint damping parameters that allow for stable long-term evolutions while
minimally decreasing the accuracy of the solution. For single neutron star
simulations we use
\begin{eqnarray}
  \label{eq:gamma0 single ns}
  \gamma_0 &= 0.12 \exp\left(-\frac{r^2}{7.884855^2}\right) + 0.01, \\
  \label{eq:gamma1 single ns}
  \gamma_1 &= 0.999 \exp\left(-\frac{r^2}{30^2}\right) - 0.999, \\
  \label{eq:gamma2 single ns}
  \gamma_2 &= 1.2 \exp\left(-\frac{r^2}{7.884855^2}\right) + 0.01,
\end{eqnarray}
where $r$ is the coordinate radius $r=\sqrt{x^i\delta_{ij}x^j}$ with
$\delta_{ij}$ the Kronecker delta symbol. For binary neutron star mergers we use
\begin{eqnarray}
  \gamma_i &=
  \gamma_{i,A} \exp\left[\frac{|x^j-x^j_A|^2}{w_A^2}\right] +
  \gamma_{i,B} \exp\left[\frac{|x^j-x^j_B|^2}{w_B^2}\right] \nonumber \\
  &+ \gamma_{i,C} \exp\left[\frac{r^2}{w_C^2}\right] +
  \gamma_{i,D}
\end{eqnarray}
where $x^i$ is the grid-frame coordinates, $x^i_{A,B}$ are the locations of the
centers of the neutron stars in the grid frame, and the other parameters are
freely specifiable constant. For the three parameters $\gamma_{0,1,2}$ entering
the GH equations, we use
\begin{eqnarray}
  \gamma_{0,A}=\gamma_{0,B}= \gamma_{0,C} = 0.06277857994;\,\, \gamma_{0,D}=0.01;\\
  \gamma_{1,C}=0.999;\,\, \gamma_{1,A}=\gamma_{1,B}=0;\,\, \gamma_{1,D}=-0.999;\\
  \gamma_{2,A}=\gamma_{2,B} = 0.94167869922;\,\, \gamma_{2,C} = 0.19182343873;\,\, \gamma_{0,D}=0.01;\\
  w_A = w_B = 7.884855;\,\, w_C = 51.60996.
\end{eqnarray}
These choices are drawn from our experience running similar systems with
\texttt{SpEC} where we use the below functional forms depending on the masses of
the stars $M_A$ and $M_B$ in solar masses. For $\gamma_0$ we use
\begin{eqnarray}
  \gamma_{0,A} = \frac{0.09}{M_A},\;\;
  \gamma_{0,B} = \frac{0.09}{M_B},\;\;
  \gamma_{0,C} = \frac{0.18}{M_A + M_B},\;\;
  \gamma_{0,D} = 0.01.
\end{eqnarray}
For $\gamma_1$ we use
\begin{eqnarray}
  \gamma_{1,A} = 0,\;\;
  \gamma_{1,B} = 0,\;\;
  \gamma_{1,C} = 0.999,\;\;
  \gamma_{1,D} = -0.999,
\end{eqnarray}
which makes the zero-speed constraint fields propagate radially outward at the
outer boundary. For $\gamma_2$ we use
\begin{eqnarray}
  \gamma_{2,A} = \frac{1.35}{M_A},\;\;
  \gamma_{2,B} = \frac{1.35}{M_B},\;\;
  \gamma_{2,C} = \frac{0.55}{M_A + M_B},\;\;
  \gamma_{2,D} = 0.01.
\end{eqnarray}
Finally, the weights are given by
\begin{eqnarray}
  w_{A} = 5.5 M_A,\;\;
  w_{B} = 5.5 M_B,\;\;
  w_{C} = 18 (M_A + M_B).
\end{eqnarray}

\section{Numerical results\label{sec:dgscl numerical results}}
We now present numerical results from single and binary neutron star
simulations. We refer to an element using $(N+1)^3$ DG points as a $P_N$
element or as using a $P_N$ scheme. The corresponding FD grid has $(2N+1)^3$ FD
grid points.

\subsection{Single Star}

We begin our numerical evaluation of \texttt{SpECTRE}'s DG-FD hybrid method by
simulating several configurations of single stars in equilibrium in full 3d
using the harmonic gauge $H_a=0$. We use the \blue{Harten-Lax-van Leer (HLL)}
Riemann solver~\cite{HLL} on all elements.
Time evolution is performed using a third-order Runge-Kutta method. In all
tests, the domain consists of an inner cube covering the region $[-17.8,17.8]^3$
with a transition layer to a spherical boundary at $r=100$ and a surrounding
spherical shell covering $r\in[100, 250]$. The transition layer and spherical
shell are divided into $6$ regions with $90^\circ$ opening angles (i.e.~a cubed
sphere geometry). We vary the resolution across the scenarios, but by
convention, the cube consists of $K^3$ elements, where we may have
$K\in[8,16,32]$. Each region of the inner spherical shells consists of $(K/2)^3$
elements, and each region the outer shells consists of
$(K/2)_r\times(K/4)_{\theta,\phi}^2$ elements. For each value of $K$, we may
further vary the resolution by changing the number of basis functions used by
each DG element. Specifically, we use $P_5$ through $P_7$ elements. This choice
is uniform across the entire domain, except when $K=32$, in which case the
shells only use $P_5$ elements, even if the central cube uses $P_6$ or $P_7$
ones.

All single star simulations are evolved with a polytropic equation of state,
\begin{equation}
  p(\rho)=\kappa\rho^\Gamma,
\end{equation}
with the polytropic exponent $\Gamma=2$, polytropic constant $\kappa=100$.

The following subsections outline the various tests that were run adopting this
setup.

\begin{table}[]
  \hfill
  \begin{tabular}{l|lll|lll|lll}
\hline
$K$                &       & $8$   &       &       & $16$  &       &       & $32$  &       \\
$P_i$              & $P_5$ & $P_6$ & $P_7$ & $P_5$ & $P_6$ & $P_7$ & $P_5$ & $P_6$ & $P_7$ \\ \hline
$\Delta x$ {[}m{]} &    599   &   507   &    439   &    299   &    253   &    219   &   150    &      127  &   110 \\ \hline
\end{tabular}
  \caption{Resolutions of the FD grid in the central cube corresponding to our
    chosen grid parameters for the single star runs. The star is located within
    this central cube, which consists of $K^3$ $P_i$ elements.}
  \label{table:resolution}
\end{table}

\subsubsection{TOV star}

In the case of a static, spherically symmetric star, we follow the procedures
of \cite{Deppe:2021bhi, Deppe:2021ada}. Namely, we construct a star using the
Tolman-Oppenheimer-Volkoff (TOV) solution \cite{Tolman:1939jz,
Oppenheimer:1939ne}. The star's central density is $\rho_c=1.28\times10^{-3}$,
such that the total mass in this solution is $M=1.4M_\odot$. For FD cells in
these simulations, we use the monotonized central reconstruction
method~\cite{VanLeer:1977}. We use this case to provide an in-depth exploration
of the effects of resolution on our results. As such, we test all
$K\in[8,16,32]$ and all DG basis functions $P_5$ through $P_7$. In all cases,
we run the simulation to $t=10\,{\rm ms}$.

\begin{figure}
  \includegraphics[width=0.513\columnwidth]{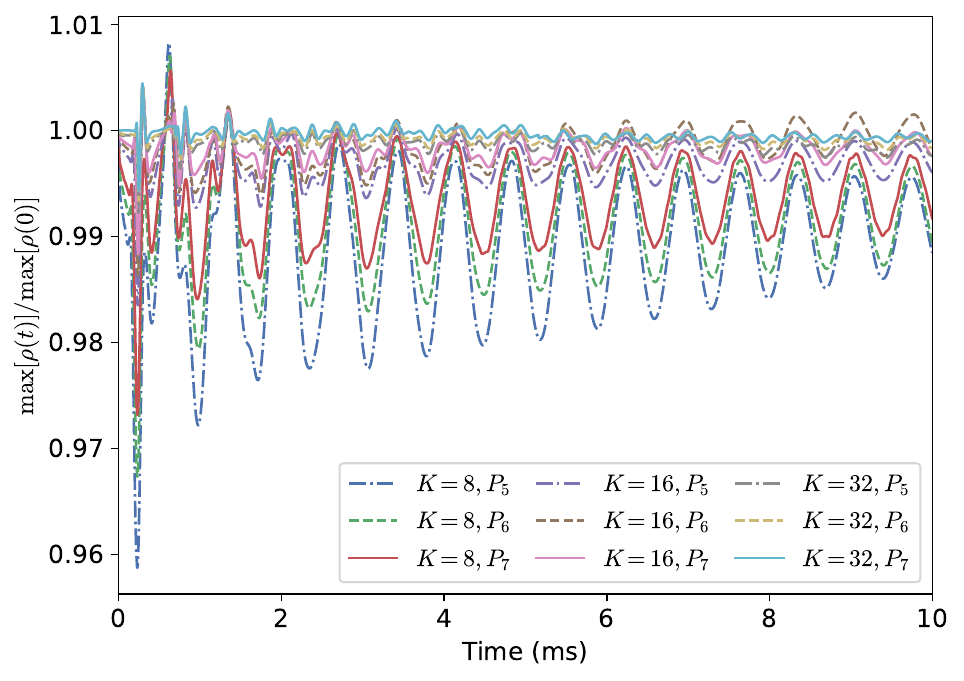}
  \includegraphics[width=0.487\columnwidth]{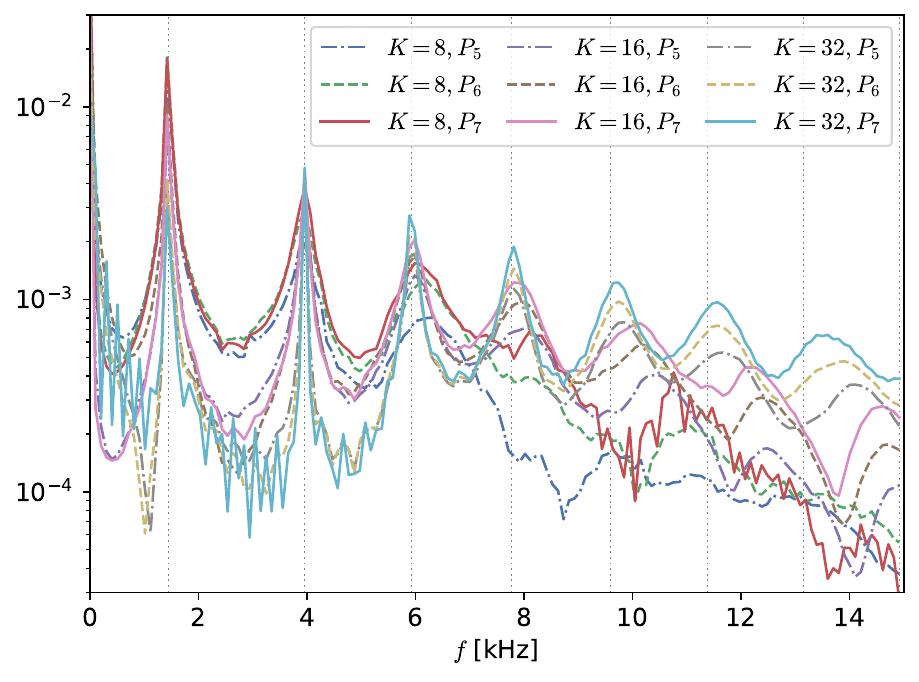}
  \caption{{\it Left}: Maximum value of the baryon rest mass density on the
    computational domain for all TOV star simulations, normalized to the maximum
    rest mass density at $t=0$. {\it Right}: The power spectra of the maximum
    rest mass density for all the above cases. Vertical dashed lines correspond
    to the known dynamical spacetime frequencies for the system.}
  \label{fig:tov-maxrho-spectrum}
\end{figure}

The left panel of figure~\ref{fig:tov-maxrho-spectrum} shows the evolution of
the normalized maximum rest mass density over time. The $K=8$ and $K=16$
simulations use FD throughout the entire stellar interior, and although the
$K=32$ simulations use DG cells for at least part of the stellar interior, the
$P_5$ and $P_6$ cases still use FD cells at the stellar core. This means that
all simulations except for $K=32$, $P_7$ will not have a grid point at the
center and will have an offset in central density from the initial value.
However, as the resolution improves, the density converges toward the initial
value. The right panel of figure~\ref{fig:tov-maxrho-spectrum} shows the same
data in the frequency domain, demonstrating that the oscillations in the data
can largely be attributed to the known radial oscillation
modes~\cite{2002PhRvD..65h4024F}. We find that even the lowest resolution
simulation resolves two of the modes well, and as resolution improves, so does
the quality and quantity of resolved modes.

\subsubsection{Rotating neutron star}

To generate initial data for a uniformly rotating star, we numerically solve the
Einstein-hydrostatic equilibrium equations according to the methods
of~\cite{1992ApJ...398..203C, 1994ApJ...424..823C}. In our case, we generate a
star of gravitational mass $M=1.627M_\odot$ and rotational period
$1.820\,{\rm ms}$, such that the ratio of polar to equatorial radius is
$0.85$. We then load this initial data into \texttt{SpECTRE} and evolve the
system. For FD cells in these simulations, we use the PPAO reconstruction
method~\cite{Deppe:2023qxa}.  We test this scenario at two resolutions,
$K\in[16,32]$, both using the $P_6$ scheme, and we run both simulations to
$t=5\,{\rm ms}$.

\begin{figure}
  \hfill
  \includegraphics[width=0.513\columnwidth]{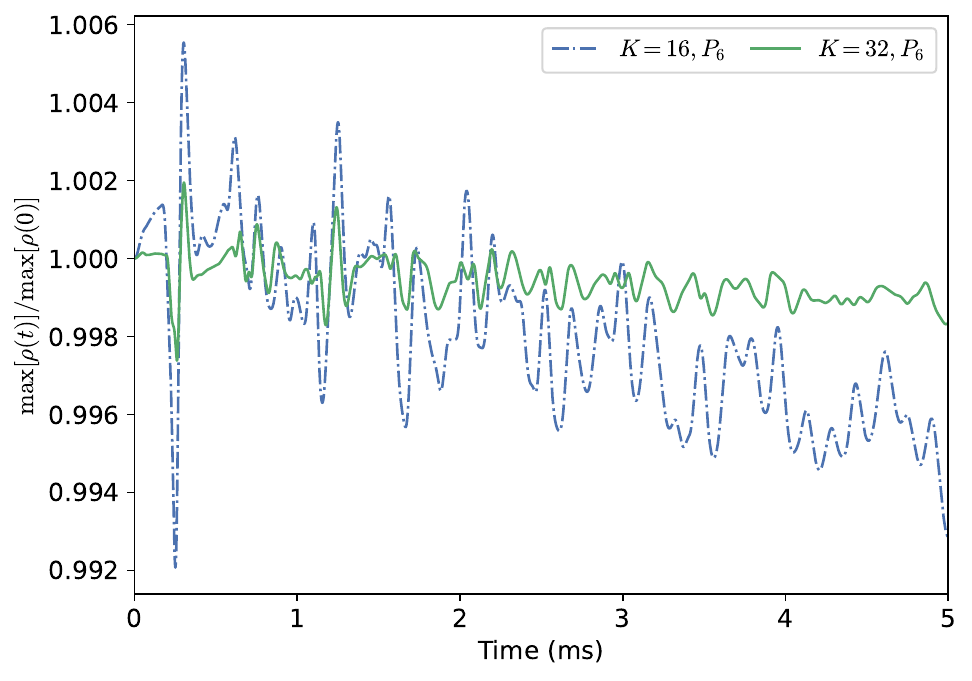}
  \caption{Maximum value of the baryon rest mass density on the
  computational domain for all Rotating star simulations, normalized to the
  maximum rest mass density at $t=0$.}
  \label{fig:rot-maxrho-spectrum}
\end{figure}

As with the TOV case, figure~\ref{fig:rot-maxrho-spectrum} depicts the evolution
of the maximum rest mass density over time.  We see a decay in this maximum
density over time, which is a known consequence of the dissipative nature of FD
schemes~\cite{Deppe:2021ada}. For the lower resolution $K=16$ case, the decay is
sub-percent level over our considered duration, and increasing the simulation
resolution further reduces its effect.

\subsection{Binary neutron star merger\label{sec:bns mergers}}

\subsubsection{Numerical setup}

To test \texttt{SpECTRE}'s ability to evolve binary neutron star (BNS) systems
with the DG-FD hybrid method, we perform a series of BNS simulations with both
\texttt{SpECTRE} and \texttt{SpEC}. For ease of comparison the simulations use
the same initial conditions, gauge choice, and constraint damping
parameter. Specifically, we consider an equal mass system with neutron stars of
gravitational masses $M_A=M_B=1.35M_\odot$. The stars have initial separation
$d_0=47\,{\rm km}$ in the coordinates of our initial data. The center of the
neutron stars are initially at $x^i_A=(16,0,0)M_\odot$,
$x^i_B=(-16,0,0)M_\odot$, with initial angular velocity $\Omega_0 (M_A+M_B)=0.0223$. The
neutron stars have coordinate radii $R=9.48$ ($14\,{\rm km}$).

We generate quasi-circular data with the \texttt{Spells} elliptic
solver~\cite{Pfeiffer2003,FoucartEtAl:2008,Haas:2016}. We use the simple ideal
gas equation of state
\begin{eqnarray}
  p &= \kappa \rho^\Gamma + \rho T
      =\kappa\rho^\Gamma+\rho
      (\Gamma-1)\left(\epsilon-\frac{K\rho^{\Gamma-1}}{\Gamma-1}\right), \\
  T&=(\Gamma-1)\left(\epsilon-\frac{K\rho^{\Gamma-1}}{\Gamma-1}\right), \\
  \epsilon &= \frac{1}{(\Gamma-1)}\frac{p}{\rho}
\end{eqnarray}
with $p$ the pressure, $\rho$ the baryon density, $\epsilon$ the specific
internal energy, and $T$ a ``temperature'' variable effectively parametrizing
the thermal energy of the fluid. We use $\Gamma=2$ and $\kappa=123.6$. All
evolutions are performed using the harmonic gauge $H_a=0$.

Besides these choices, the \texttt{SpEC} and \texttt{SpECTRE} simulations use
very different numerical methods. The \texttt{SpEC} simulations use the standard
setup described in~\cite{Duez:2008rb,Foucart:2013a,Haas:2016}, i.e.~a
pseudospectral grid for the evolution of the GH equations made of a small number
of large subdomains adapted to the geometry of the system (78 pseudospectral
subdomains during inspiral, including spheres close to the compact objects and
in the wave zone, distorted cylinders and blocks in between), finite volume
evolution of the hydrodynamics equations in the Valencia
formalism~\cite{2006ApJ...637..296A} with fifth-order \blue{weighted essentially non-oscillatory
(WENO5)}~\cite{Liu1994200,Jiang1996202,Borges} reconstruction from cell
centers to cell faces and an HLL~\cite{HLL} approximate Riemann solver to
calculate numerical fluxes on those faces. The finite difference grid is itself split
during inspiral into $\sim (400-700)$ elements, with the number of elements
changing over time as matter covers an increasingly large fraction of the
pseudospectral grid. For the low-resolution \texttt{SpEC} simulation, each FD
element uses $16^3$ grid points excluding ghost zones and for the
medium-resolution simulation each FD element uses $19^3$ grid points excluding
ghost zones. Ghost zones add 6 grid points per dimension for each
simulation. Time evolution is performed using a third-order Runge-Kutta
method. We run the \texttt{SpEC} simulations at 2 resolutions corresponding to
our standard ``low'' and ``medium'' resolution settings
($\Delta x =[0.263,0.211]M_\odot$ on the finite difference grid). In
\texttt{SpEC}, the grid both rotates and contracts with the evolution of the
neutron stars, in such a way as to keep the center of mass of each star fixed on
the numerical grid.

\begin{figure}
  \includegraphics[width=0.5\columnwidth]{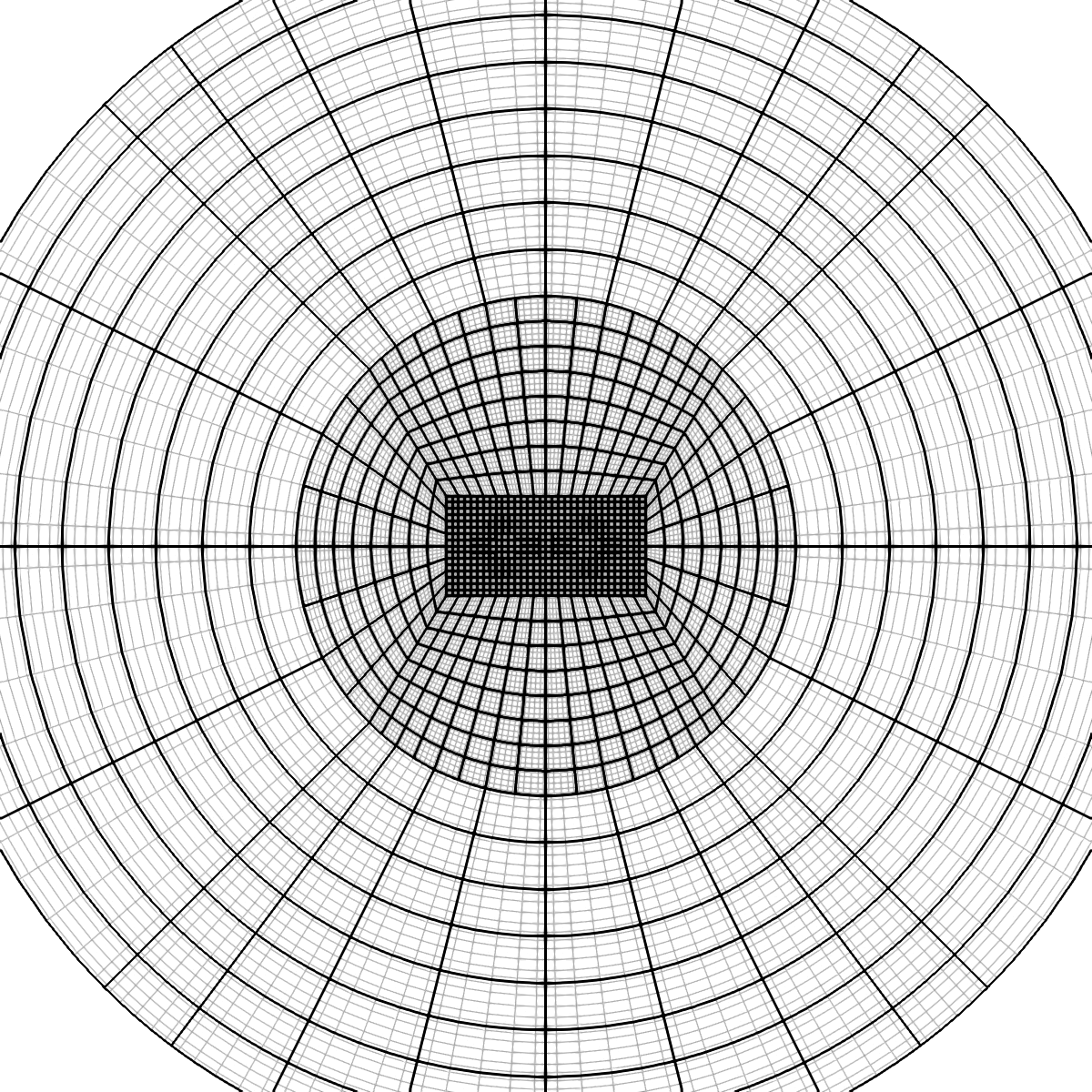}
  \includegraphics[width=0.5\columnwidth]{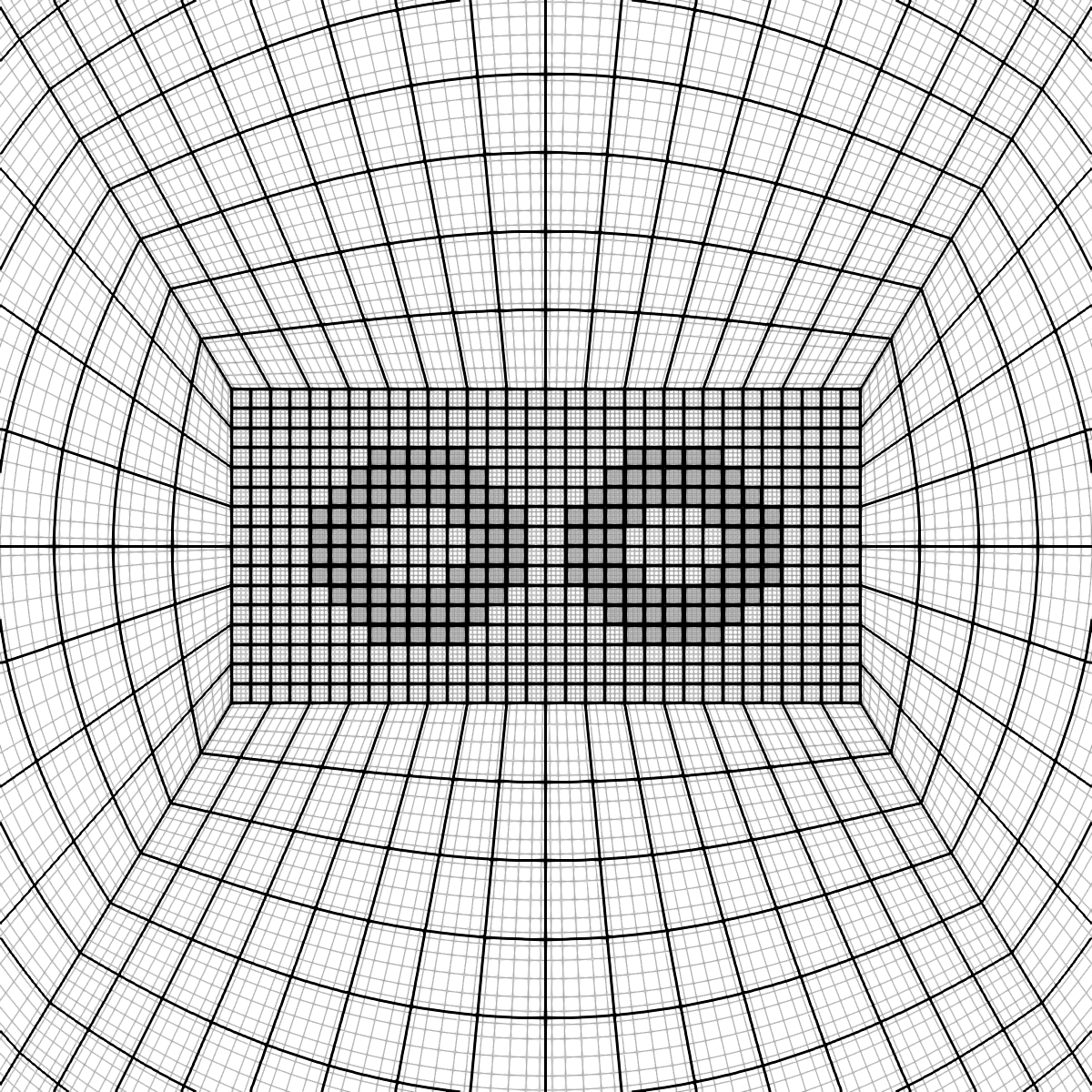}
  \caption{{\it Left}: Domain decomposition for the \texttt{SpECTRE}
    simulations. The central rectangle covers the two orbiting neutron stars,
    while curved elements extend from that box into the wave zone. Black lines
    denote element boundaries while intersections of grey lines are the grid
    point locations. {\it Right}: A zoom in on the central rectangle. FD is used
    in the circular regions with denser grey lines, which tracks the stellar
    surfaces. Note that which elements use FD or DG changes dynamically during
    simulations based on our troubled-cell indicator (TCI).}
  \label{fig:domain}
\end{figure}

\begin{figure}
  \hfill
  \includegraphics[width=0.83\columnwidth]{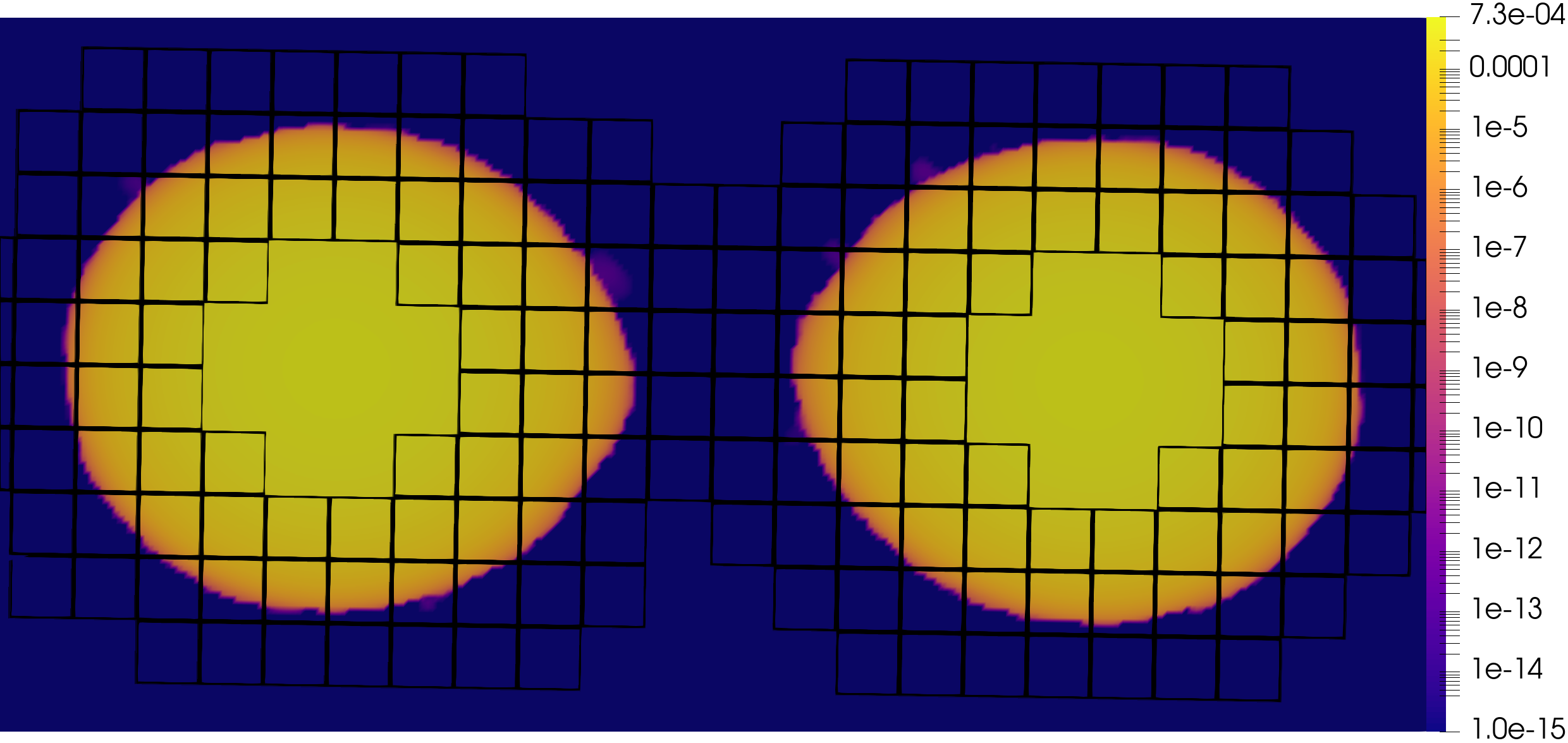}
  \caption{Baryon rest mass density at about $t=3.5\,{\rm ms}$ from the $P_6$
    \texttt{SpECTRE} simulation. The elements outlined in black use FD. We see
    that during inspiral \texttt{SpECTRE} uses FD methods close to the stellar
    surfaces, and DG methods elsewhere.}
  \label{fig:bns rest mass density}
\end{figure}

\texttt{SpECTRE} on the other hand uses a larger number of smaller elements
(14592 elements in total), making use of the mixed DG-FD algorithm described
above. On FD elements, we use the PPAO reconstruction method\cite{Deppe:2023qxa}
and HLL Riemann solver. Time evolution is performed using a third-order
Adams-Bashforth method~\cite{BA66330495}. We show the \texttt{SpECTRE} domain
decomposition in figure~\ref{fig:domain}. The inner part of the domain is
constructed from a grid of $32\times 16\times 16$ cubes covering the region
$[-40,40]\times [-20,20]\times [-20,20]$ around the neutron stars. The outer
part of the domain is a shell covering radii $r\in [100,250]$, divided into $6$
regions with $90^\circ$ opening angles (``JuggleBalls'' geometry). Each of these
regions is divided into $8\times 4\times 4$ cubes (8 in the radial
direction). Finally, the inner and outer region are connected by an envelope of
10 distorted cubes, each divided into $8\times 8\times 8$ elements. We vary
resolution by changing the number of basis functions used by each DG
elements. Specifically, we use $P_4$ through $P_7$ elements. Around the neutron
stars, and when using FD, our effective grid resolution is thus
$\Delta x =[0.278,0.227,0.192,0.167]$. Practically, our experimentation with a
range of different domain decompositions indicate that for $P_4$, our errors are
dominated by inaccurate evolution of the GH equations in the envelope and outer
region. This is not surprising given the GH solver is running at only fourth
order in this case. We will see that we observe a significant decrease in
numerical error for e.g., the trajectories of the neutron stars, between $P_4$
and $P_5$. In \texttt{SpECTRE}, the computational domain rotates with the
neutron stars, but we do not perform any rescaling, i.e.~the center of mass of
the two neutron stars remains along the $x$-axis in grid coordinates, but the
stars approach each other on that axis over the course of the evolution.  In
figure~\ref{fig:bns rest mass density} we plot the baryon rest mass density at
$t\approx3.5$ms. The elements outlined in black use FD. We see that during
inspiral \texttt{SpECTRE} uses FD methods close to the stellar surfaces and DG
methods elsewhere. During the merger itself, larger fractions of the
computational domain switch to FD; how to optimize when to switch between FD and
DG methods during the post-merger phase remains an open question.

\texttt{SpEC} and \texttt{SpECTRE} use the same mechanism to ``correct'' the
evolved variables in low-density regions~\cite{Foucart:2013a,
  Muhlberger2014}. While \texttt{SpEC} additionally corrects the primitive
variables (temperature, velocity) at densities
$\rho<2\times 10^{-11}$ (i.e.~roughly $8$ orders of magnitude below
the central density of the neutron star), \texttt{SpECTRE} only sets velocities
to small ($<10^{-4}$) values at $\rho<9\times 10^{-15}$, and does
not correct the temperature. \texttt{SpEC} uses a density floor of
$\rho=10^{-13}$, while \texttt{SpECTRE} uses a floor of
$\rho=10^{-15}$.

\subsubsection{Results}

\begin{figure}
  \includegraphics[width=0.5\columnwidth]{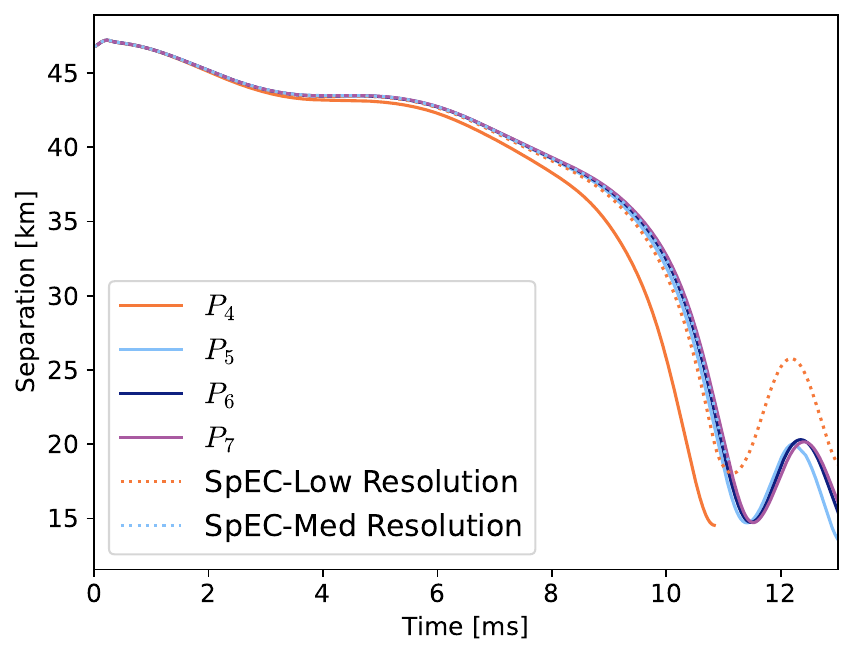}
  \includegraphics[width=0.5\columnwidth]{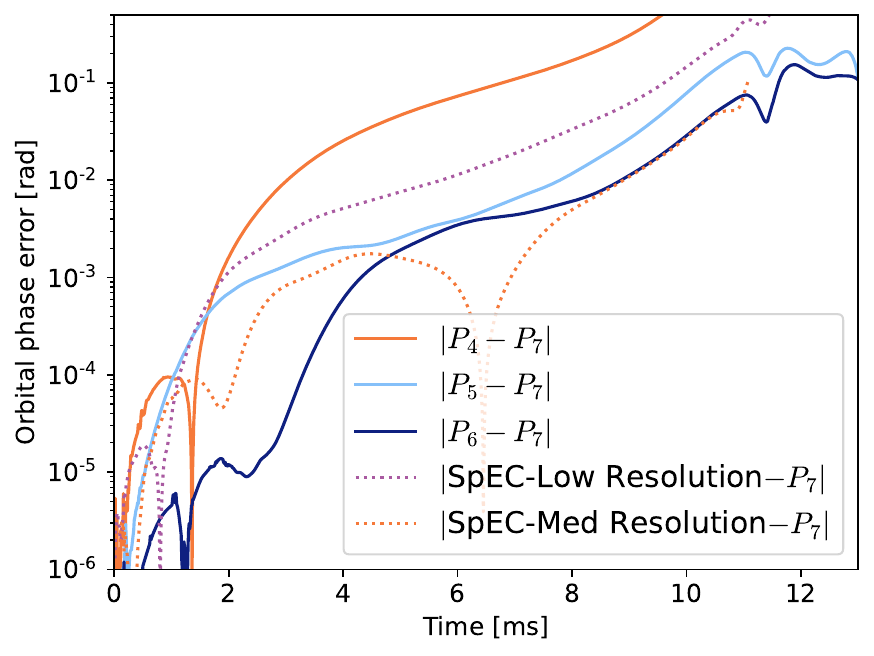}
  \caption{{\it Left}: Binary separation as a function of time for all six
    simulations (the MR \texttt{SpEC} simulation largely tracks the $N=7$
    \texttt{SpECTRE} simulation). {\it Right}: Difference in orbital phase with
    respect to the highest resolution \texttt{SpECTRE} simulation, for the
    $N=6,7$ \texttt{SpECTRE} simulations and the two \texttt{SpEC} simulations.}
  \label{fig:bns-traj-conv}
\end{figure}

Each neutron star goes through about $3.5$ orbits
before merger, on slightly eccentric orbits. Figure~\ref{fig:bns-traj-conv}
provide us with more information about the numerical accuracy of the
simulations. In the left panel, we show the binary separation as a function of
time for all simulations. As previously mentioned, the $P_4$ \texttt{SpECTRE}
simulation is significantly less accurate than all other simulations, likely
because of the low-order methods used in the envelope. Experimentally we found
that the $P_4$ scheme is fairly sensitive to the choice of domain decomposition
in the outer regions. The $P_5$ through $P_7$ \texttt{SpECTRE} simulations show
clear convergence of the merger time, with the \texttt{SpEC} and
\texttt{SpECTRE} simulations agreeing within estimated numerical error. In the
right panel we show the phase error, estimated here as the orbital phase
difference with the $P_7$ \texttt{SpECTRE} simulation.
The simulations show both the \texttt{SpEC} and \texttt{SpECTRE}
simulations quickly approaching the results of the high-resolution
\texttt{SpECTRE} simulations as resolution increases. We note that the
trajectory and phase do not show clean pointwise convergence at all times, due
to crossings in the trajectories at the time of periastron passage; i.e.~around
$4-5\,{\rm ms}$. This is particularly visible in the phase difference for the
\texttt{SpECTRE} simulations at time $\sim 5\,{\rm ms}$.

\begin{figure}
  \includegraphics[width=0.5\columnwidth]{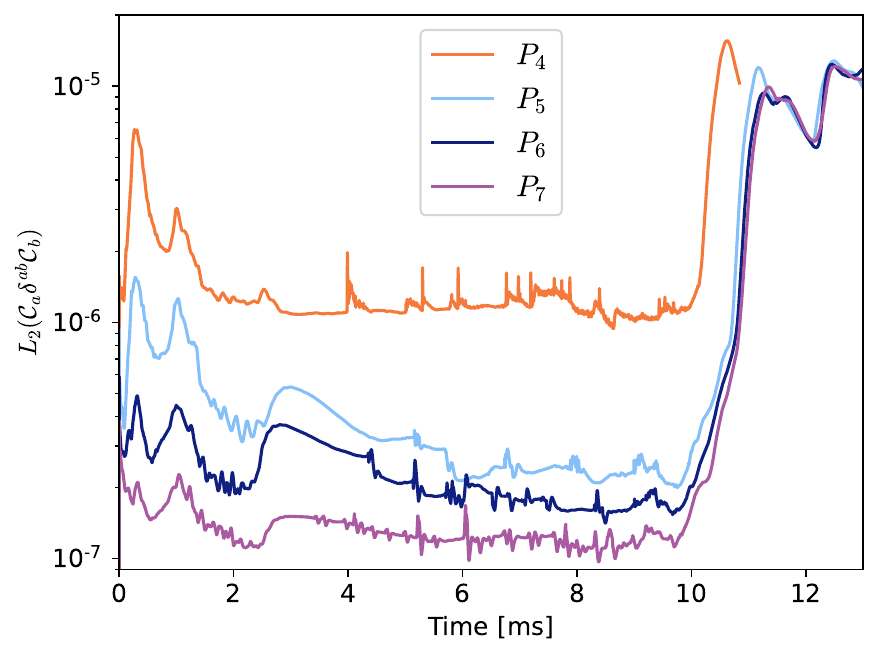}
  \includegraphics[width=0.5\columnwidth]{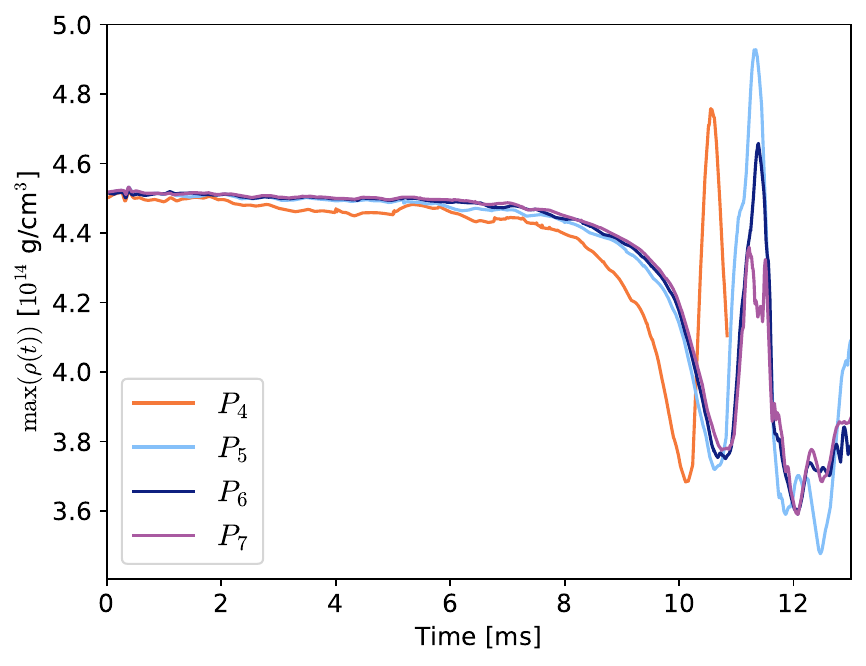}
  \caption{{\it Left}: Violation of the gauge constraint integrated over the
    computational domain, for all \texttt{SpECTRE} simulations. {\it Right}:
    Maximum value of the baryon density on the computational domain, for all
    \texttt{SpECTRE} simulations.}
  \label{fig:bns-cons-maxrho}
\end{figure}

In figure~\ref{fig:bns-cons-maxrho}, we show quantities that more directly track
the error in the evolution of the GH and hydrodynamics equations. The left panel
shows the violation of the gauge constraint, integrated over the entire
computational domain. We see clear convergence with resolution before merger. At
merger, we do not necessarily expect convergence as we introduce a fixed source
of error at the boundary of the region where we allow matter to evolve. In the
right panel we plot the maximum value of the baryon density on the computational
domain. Errors in the evolution of the fluid equations typically lead to a slow
decrease in the value of maximum density during inspiral---in addition to the
physically expected decrease of the maximum density as each star is tidally
distorted by its companion. We indeed see higher dissipation for $P_4$, and
convergence of the evolution of the maximum density for $P_5, P_6, P_7$ up to
contact, at least in a time-averaged sense (i.e.~ignoring the oscillations of
the stars that do not remain exactly in phase at all resolutions, and lead to
out-of-phase oscillations of the central density).

\begin{figure}
  \hfill
  \includegraphics[width=0.93\columnwidth]{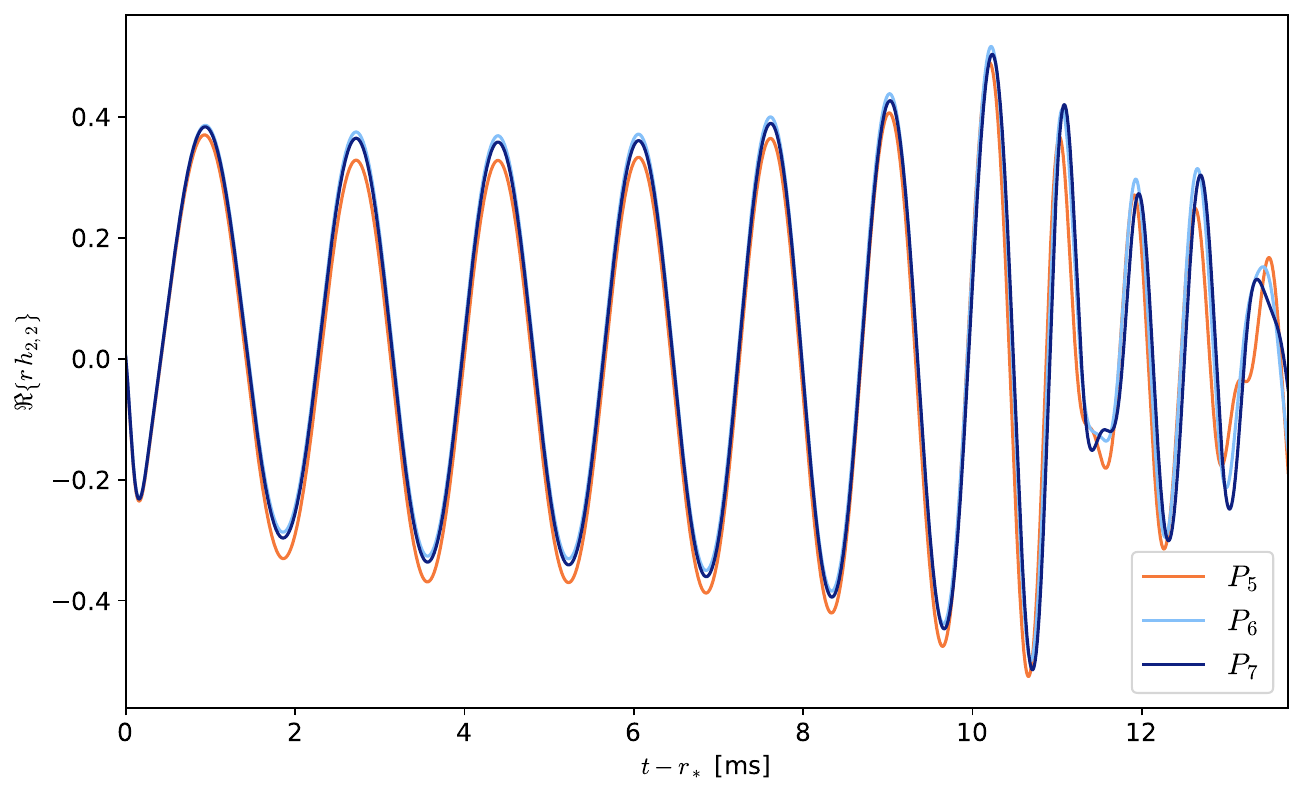}
  \caption{Shown is the real part the $(2,2)$ of the strain $rh$ for the binary
    merger simulation at three different resolutions. The waveform is extracted
    using \texttt{SpECTRE}'s Cauchy-Characteristic Evolution
    capabilities. Convergence with increasing resolution of the Cauchy evolution
    is observed. \blue{That is, the $P_7$ and $P_6$ simulations are closer
      together than the $P_6$ and $P_5$ simulations. A careful study of the
      convergence order and how it depends on different choices in the methods
      will be presented in a followup paper.}}
  \label{fig:bns waveforms}
\end{figure}

Finally, we show the first binary neutron star gravitational waveforms extracted
using Cauchy-Characteristic Evolution (CCE)~\cite{2020PhRvD.102d4052M,
  Moxon:2021gbv, 1996PhRvD..54.6153B, 1999bhgr.conf..383B, 2005CQGra..22.2393B,
  2015CQGra..32b5008H, 2015CQGra..32w5018H, 2016LRR....19....2B,
  2016CQGra..33p5001C, 2016CQGra..33v5007H, 2020PhRvD.102b4004B}. In
figure~\ref{fig:bns waveforms} we plot the real part of the $(2,2)$ mode of the
strain $h$ using an extraction worldtube located at $r=200M$ at the three
highest resolutions. We see convergence with increasing resolution of the Cauchy
evolution.  \blue{That is, the $P_7$ and $P_6$ simulations are closer together
  than the $P_6$ and $P_5$ simulations. A careful study of the convergence order
  and how it depends on different choices in the methods will be presented in a
  followup paper.} The CCE discretization errors are negligible compared to the
errors in Cauchy evolution.
We evolve the CCE equations in spherical coordinates, with the evolved variables
expended radially using a Legendre-Gauss-Lobatto basis and spin-weighted
spherical harmonics in the angular directions. The characteristic evolution uses
a fifth-order Adams-Bashforth local time stepper\cite{doi:10.1137/19M1292692}
with an absolute error tolerance of $10^{-8}$ and a relative tolerance of
$10^{-6}$ for the spin-weighted spherical harmonic variables and a relative
error of $10^{-7}$ for the coordinate variables. We use $l_{\max}=20$ for the
expansion in spin-weighted spherical harmonics and filter out the top 2 modes.
The radial grid uses 15 grid points and an exponential filter as in
Eq.~\ref{eq:exponential filter} with $a=35$ and $b=64$. The CCE data on the
initial slice is determined using the conformal factor
method~\cite{2020PhRvD.102d4052M, Moxon:2021gbv}. The CCE quantities like the
strain, news, and Weyl scalars are output at $\mathcal{I}^+$ up to and including
$l_{\max}=8$. CCE also needs data on a worldtube from the Cauchy evolution as
radial boundary conditions for the system. This data is kept at a resolution of
$l_{\max}=16$. We can ignore the effects of matter in our characteristic
evolution because the extraction radius is in a region of atmosphere. \blue{That
  is, the effects of matter are neglected in CCE since none of the matter
  reaches the extraction radius.} The effects of strongly gravitating matter
near the extraction radii on the characteristic evolution have not been studied
yet.  We also perform a supertranslation using the
\texttt{scri}\cite{mike_boyle_2020_4041972, PhysRevD.93.084031, Boyle:2014ioa,
  PhysRevD.87.104006} and \texttt{sxs}\cite{Boyle_The_sxs_package_2024} packages
to set the strain to zero at retarded time zero. No time or phase alignment is
done. While \texttt{SpECTRE} also outputs the news and all the Weyl scalars, we
leave a careful analysis of CCE waveforms from BNS mergers as future work,
pending a more detailed understanding of how various resolution, control system,
extraction radii, and other choices affect the accuracy and convergence of the
waveforms.

These simulations clearly demonstrate that the DG-FD hybrid scheme is capable of
accurately evolving binary neutron star systems up to and through merger, as
long as sufficiently high-order methods are used in the envelope and wave
zone. Note that \texttt{SpECTRE} and \texttt{SpEC} have distinct (dis)advantages
in the context of binary neutron star evolutions. \texttt{SpEC} allows for very
cost-effective evolutions of the binary thanks to spectral domains adapted to
the geometry of the system. However, the use of a small number of large elements
prevents \texttt{SpEC} from scaling beyond $\mathcal{O}(100)$ processors at low
resolution. Additionally, attempting to go to higher resolution with
\texttt{SpEC} leads to simulations whose cost is dominated by a few elements
with large numbers of basis functions, typically situated around the surfaces of
the stars. This would lead to extremely poor load-balancing, and has thus far
limited our ability to perform higher resolution simulations. \texttt{SpECTRE},
on the other hand, is less cost-effective at low resolutions but can leverage a
higher number of processors and, by using FD methods close to the surface of the
star, is less sensitive to high-frequency noise in that region. In the
simulations presented here, which use \texttt{SpEC} domains optimized over more
than 10 years of experimentation but very unoptimized \texttt{SpECTRE} domain
decompositions, the higher resolution \texttt{SpEC} simulation cost $14.2k$
CPU-hrs on the Wheeler cluster at Caltech to reach $t=11\,{\rm ms}$. Wheeler has
two 12-core Intel\textsuperscript{\tiny{\textregistered}}
Xeon\textsuperscript{\tiny{\textregistered}} CPU E5-2680 v3 with a base clock of
2.50GHz CPUs per node. The $P_5$ \texttt{SpECTRE} simulation, with a FD grid
spacing $5\%$ coarser at the location of the neutron stars, costs $27.8k$
CPU-hrs, but used $288$ cores instead of $120$. The $P_7$ \texttt{SpECTRE}
simulation used $117k$ CPU-hrs. This cost increase can be compared to the
expected scaling for finite difference method (with cost
$\propto \Delta x^{-4}$), which would predict a cost of $95k$ CPU-hrs for the
$P_7$ simulation. Similarly, the $P_6$ simulation cost $52k$ CPU-hrs to reach
the same time, while scaling as $\propto \Delta x^{-4}$ the $P_6$ cost predicts
a cost of $54k$ CPU-hrs. These are already promising numbers that we expect will
improve with better parallelization of the DG-FD hybrid algorithm. Specifically,
we expect on-the-fly redistribution of elements to different processes (since FD
elements are significantly more expensive than DG elements) and more optimized
domain decompositions to significantly reduce the cost. Whether \texttt{SpECTRE}
is able to outperform \texttt{SpEC}'s CPU time is currently unclear. However,
one \texttt{SpECTRE}'s primary goals is to reduce wall time by scaling to more
processors.

\section{Conclusions\label{sec:dgscl conclusions}}
In this paper we gave a detailed description of our DG-FD hybrid method that can
successfully solve challenging general relativistic astrophysics problems in
dynamical spacetimes, including the simulation of a neutron star, a rotating
neutron star, and a binary neutron star merger. Our method combines an unlimited
DG solver with a conservative FD solver. Alternatively, this can be thought of
as taking a standard FD code in numerical relativity and compressing the data to
a DG grid wherever the solution is smooth. The DG solver is more efficient than
the FD solver since no reconstruction is necessary and fewer Riemann problems
need to be solved. The algorithm presented here is an extension of our previous
work in static spacetimes~\cite{Deppe:2021ada}. The basic idea is that an
unlimited DG solver is used wherever a troubled-cell indicator deems the DG
solution admissible, while a FD solver is used elsewhere. Unlike classical
limiting strategies like WENO~\cite{Zhong2013397, 2016CCoPh..19..944Z} which
attempt to filter out unphysical oscillations, the hybrid scheme prevents
spurious oscillations from entering the solution. This is achieved by retaking
any time step using a robust high-resolution shock-capturing conservative FD
scheme where the DG solution was inadmissible, either because the DG scheme
produced unphysical results like negative densities, or because a numerical
criterion like the percentage of power in the highest modes deemed the DG
solution bad. Our DG-FD hybrid scheme was used to perform the first
simulations of a rotating neutron star and binary neutron star merger using DG
methods. We show the first gravitational waveforms obtained from binary neutron
star mergers using Cauchy-Characteristic Evolution~\cite{2020PhRvD.102d4052M,
  Moxon:2021gbv, 1996PhRvD..54.6153B, 1999bhgr.conf..383B, 2005CQGra..22.2393B,
  2015CQGra..32b5008H, 2015CQGra..32w5018H, 2016LRR....19....2B,
  2016CQGra..33p5001C, 2016CQGra..33v5007H, 2020PhRvD.102b4004B}, though leave a
detailed analysis of the waveforms to future work\footnote{Cauchy-Characteristic
  Evolution currently does not take the effects of matter passing through the
  worldtube into account and so long-term post merger wave extraction will
  require careful study.}. In the future we plan to improve our handling of
curved meshes to allow tracking outflows in the post-merger phase, incorporate
constrained transport for ensuring $\partial_i B^i=0$~\cite{Londrillo:2003qi,
  Gardiner:2007nc, Cunningham:2007mn, Miniati:2011ww, Mignone:2011fd,
  Lee:2013xfa, Balsara:2017kyz, Mignone:2020qec, Kiuchi:2022ubj, Seo:2023yra,
  Berta:2023zmr}, use local adaptive time stepping using a linear multi-step
method~\cite{doi:10.1137/19M1292692}, use adaptive mesh refinement
(e.g.~\cite{Szilagyi:2014fna, Schaal:2015ila, Renkhoff:2023nfw}), dynamic
continuous load-balancing, and an optimized domain decomposition.

\ack

Charm++/Converse~\cite{laxmikant_kale_2020_3972617} was developed by the
Parallel Programming Laboratory in the Department of Computer Science at the
University of Illinois at Urbana-Champaign. The figures in this article were
produced with \texttt{matplotlib}~\cite{Hunter:2007,
  thomas_a_caswell_2020_3948793}, \texttt{TikZ}~\cite{tikz} and
\texttt{ParaView}~\cite{paraview, paraview2}. Computations were performed with
the Wheeler cluster at Caltech and the mbot cluster at Cornell. This work was
supported in part by the Sherman Fairchild Foundation and by NSF Grants
No.~PHY-2309211, No.~PHY-2309231, and No.~OAC-2209656 at Caltech, and NSF Grants
No.~PHY-2207342, No.~PHY-2407742 and No.~OAC-2209655 at Cornell.  F.F. gratefully acknowledges
support from the Department of Energy, Office of Science, Office of Nuclear
Physics, under contract number DE-AC02-05CH11231, from NASA through grant
80NSSC22K0719, and from the NSF through grant AST-2107932. M.D. gratefully
acknowledges support from the NSF through grant PHY-2110287 and support from
NASA through grant 80NSSC22K0719. GL and MSB acknowledge support from NSF award
PHY-2208014, the Dan Black Family Trust and Nicholas and Lee
Begovich. I.L. acknowledges support from the Department of Energy under award
number DE-SC0023101. ERM
acknowledges support by the National Science Foundation under Grant
No. AST-2307394 and PHY-2309210, the NSF Frontera supercomputer under grant
AST21006, and Delta at the National Center for Supercomputing Applications
(NCSA) through allocation PHY210074 from the Advanced Cyberinfrastructure
Coordination Ecosystem: Services \& Support (ACCESS) program, which is supported
by National Science Foundation grants \#2138259, \#2138286, \#2138307,
\#2137603, and \#2138296. ERM further acknowledges support on Perlmutter through
NERSC under grant m4575.  P.K. acknowledges support of the Department of Atomic
Energy, Government of India, under project no.~RTI4001, and by the Ashok and
Gita Vaish Early Career Faculty Fellowship at the International Centre for
Theoretical Sciences.

\newcommand\aap{Astron.~Astrophys.~}
\newcommand\mnras{Mon.~Not.~R.~Astron.~Soc.~}
\newcommand\prd{Phys.~Rev.~D }

\section*{References}
\bibliographystyle{unsrt}
\bibliography{refs}
\end{document}